\documentclass[fleqn,usenatbib]{mnras}

\usepackage[T1]{fontenc}
\usepackage{ae,aecompl}

\usepackage{graphicx}	
\usepackage{amsmath}	
\usepackage{amssymb}	
\usepackage{pdflscape}
\usepackage{times,txfonts}
\usepackage{hyperref}	

\newcommand{\afe}{$\mathrm{[\alpha/Fe]}$}
\newcommand{\feh}{$\mathrm{[Fe/H]}$}

\title[Diverse $\alpha$-element abundances in galaxy discs]{The origin of diverse $\alpha$-element abundances in galaxy discs}

\author[J. T. Mackereth et al.]{
J. Ted Mackereth$^{1}$\thanks{E-mail: J.E.Mackereth@2011.ljmu.ac.uk},
Robert A. Crain$^{1}$,
Ricardo P. Schiavon$^{1}$,
Joop Schaye$^{2}$,\newauthor
Tom Theuns$^{3}$
and Matthieu Schaller$^{3}$
\\
$^{1}$Astrophysics Research Institute, Liverpool John Moores University, 146 Brownlow Hill, Liverpool L3 5RF, United Kingdom\\
$^{2}$Leiden Observatory, Leiden University, PO Box 9513, NL-2300 RA Leiden,
the Netherlands\\
$^{3}$Institute for Computational Cosmology, Department of Physics,University
of Durham, South Road, Durham DH1 3LE, UK
}

\date{Accepted XXX. Received YYY; in original form ZZZ}

\pubyear{2015}

\begin{document}
\label{firstpage}
\pagerange{\pageref{firstpage}--\pageref{lastpage}}
\maketitle

\begin{abstract}
Spectroscopic surveys of the Galaxy reveal that its disc stars exhibit a spread in \afe{} at fixed \feh{}, manifest at some locations as a bimodality. The origin of these diverse, and possibly distinct, stellar populations in the Galactic disc is not well understood. We examine the Fe and $\alpha$-element evolution of 133 Milky Way-like galaxies from the EAGLE simulation, to investigate the origin and diversity of their \afe{}-\feh{} distributions. We find that bimodal \afe{} distributions arise in galaxies whose gas accretion histories exhibit episodes of significant infall at both early and late times, with the former fostering more intense star formation than the latter. The shorter characteristic consumption timescale of gas accreted in the earlier episode suppresses its enrichment with iron synthesised by Type Ia SNe, resulting in the formation of a high-\afe{} sequence. We find that bimodality in \afe{} similar to that seen in the Galaxy is rare, appearing in approximately 5 percent of galaxies in our sample. We posit that this is a consequence of an early gas accretion episode requiring the mass accretion history of a galaxy's dark matter halo to exhibit a phase of atypically-rapid growth at early epochs. The scarcity of EAGLE galaxies exhibiting distinct sequences in the \afe{}-\feh{} plane may therefore indicate that the Milky Way's elemental abundance patterns, and its accretion history, are not representative of the broader population of $\sim L^\star$ disc galaxies.
\end{abstract}




\section{Introduction}


\label{sec:intro}
The elemental abundances of long-lived stars are a rich fossil record of the formation history of their host galaxy. Spectroscopic surveys of the Galaxy's stars have therefore long held the promise of elucidating its origin \citep[see, e.g.][and references therein]{2002ARA&A..40..487F,2013A&ARv..21...61R,2016ARA&A..54..529B}. The recent advent of surveys that measure the elemental abundances of tens to hundreds of thousands of Milky Way stars (e.g., RAVE, \citeauthor{2006AJ....132.1645S} \citeyear{2006AJ....132.1645S}; SEGUE,\citeauthor{2009AJ....137.4377Y} \citeyear{2009AJ....137.4377Y}; Gaia-ESO, \citeauthor{2012Msngr.147...25G} \citeyear{2012Msngr.147...25G}; GALAH, \citeauthor{2015MNRAS.449.2604D} \citeyear{2015MNRAS.449.2604D} and \citeauthor{2016arXiv160902822M} \citeyear{2016arXiv160902822M}; APOGEE, \citeauthor{2015arXiv150905420M} \citeyear{2015arXiv150905420M}) heralds a significant step towards realisation of the potential of what has come to be known as `galactic archaeology'. 

One of the primary  diagnostics employed by such surveys is the relationship between the abundance ratio of $\alpha$-elements and iron, \afe{}, and the iron abundance, \feh{}.  The distribution of Galactic  {\it disc} stars in the \afe{}-\feh{} plane exhibits striking trends, with observations generally revealing a spread or even bimodality in \afe{} at fixed \feh{}, both in the solar neighborhood \citep[e.g.][]{1998A&A...338..161F,2003A&A...410..527B,2000A&A...358..671G,2000AJ....120.2513P,2004AJ....128.1177V,2005A&A...433..185B,2012A&A...545A..32A,2014A&A...562A..71B} and throughout the Galactic disc \citep[][]{2014A&A...564A.115A,2014ApJ...796...38N,2015ApJ...808..132H}. These differing relative values of \afe{} are often interpreted as evidence that the populations are distinct. The population with supersolar \afe{} is generally thought to have formed rapidly, early in the history of the Galaxy, such that its progenitor gas was enriched primarily with $\alpha$-elements synthesised and promptly released by Type II supernovae (SNe), whilst incorporating relatively little iron synthesised by Type Ia SNe \citep[e.g.,][]{1989ARA&A..27..279W,1997ARA&A..35..503M}, whose contribution to the enrichment of the interstellar medium only becomes important on timescales longer than $\sim 10^9\,{\rm yr}$ after the onset of star formation.

These two components of the Galactic disc, defined purely on the basis of their abundances, are commonly associated with the structural entities referred to as the thin and thick discs \citep[e.g.][]{1998A&A...338..161F,2005A&A...433..185B,2015MNRAS.453.1855M,2016MNRAS.461.4246W}. However, when mono-abundance and mono-age populations (i.e. stars whose elemental abundances and/or ages are similar within some tolerance) are considered separately, it becomes clear that there is not a direct correlation between a star's association with the Galaxy's thick or thin disc components, and its $\alpha$-enhancement \citep[e.g.][] {2012ApJ...753..148B,2016ApJ...823...30B}, or its age \citep{2017arXiv170600018M}. Recent studies also suggest that there may be a link between the moderately metal-poor $\alpha$-enhanced  populations in the inner Galaxy and the high-\afe{} disc \citep[e.g.][]{2015A&A...577A...1D}, motivated by the fact that metal-poor, high-\afe{} stars in the Galactic centre exhibit `hotter' kinematics \citep[e.g][]{2010A&A...519A..77B,2013MNRAS.432.2092N,2014A&A...569A.103R,2016ApJ...832..132Z} than their low-$\alpha$, more metal-rich, counterparts. Although sample sizes are limited, the $\alpha$-element abundances of the bulge high-\afe{} population are similar to those of high-\afe{} stars in the local disc, at least as far as Mg and Si are concerned \citep[e.g.][]{2016AJ....151....1R,2017arXiv170202971B}. The similarities in abundances and coincident structure and kinematics between $\alpha$-enhanced bulge populations and the high-\afe{} disc may indicate a common origin of these populations \citep[e.g.][]{2010A&A...513A..35A}.

Since disc populations with markedly different \afe{} at fixed \feh{} very likely result from different star formation histories, it is challenging to reconcile their co-spatiality in the Galactic disc with the predictions of one-zone chemical evolution models \citep[e.g.,][]{1959ApJ...129..243S,1963ApJ...137..758S}. While such simple models essentially inaugurated the field of quantitative chemical evolution modeling, they had important limitations.  Chief amongst those is their failure to reproduce the metallicity distribution function (MDF) of stars in the solar neighbourhood \citep[known as the `G-dwarf problem',][]{1962AJ.....67..486V,1963ApJ...137..758S}. This motivated the development of more detailed chemical evolution models, for example, allowing for the consideration of gas inflow \citep{1972Natur.236...21L,1977ApJ...216..548T}, radial gas flow \citep[e.g.][]{2000A&A...355..929P}, and eventually the decomposition of the Galactic disc into concentric evolution zones to capture the effects of `inside out' formation \citep[e.g.][]{1976MNRAS.176...31L,1980FCPh....5..287T,1989MNRAS.239..885M}. 

Over the past two decades, two classes of analytic models in particular have attracted attention, owing to their ability to reproduce broadly the observed elemental abundance trends in the solar neighbourhood, and abundance gradients throughout the disc. These are i) models in which the high- and low-\afe{} components of the discs form in response to distinct episodes of gas accretion, during which gas is consumed at different rates \citep{1997ApJ...477..765C,2001ApJ...554.1044C,2009IAUS..254..191C}; and ii) models in which the two components represent the equilibrium star formation conditions in different parts of the disc, but later become co-incident as a consequence of the radial mixing of stars \citep{2009MNRAS.396..203S,2009MNRAS.399.1145S}. 

The `two-infall model' of \citet{1997ApJ...477..765C,2001ApJ...554.1044C}, invokes an intense initial phase of star formation whose brevity precludes enrichment of the interstellar medium (ISM) by Type Ia SNe, thus resulting in the formation of the high-\afe{} sequence. A hiatus in star formation is then invoked, during which the unconsumed fraction of the gas delivered by the first infall (which is assumed to remain in place without consumption by star formation or ejection by feedback) is enriched by Type Ia SNe, reducing its \afe{}. A second, more prolonged episode of gas infall then triggers the steady formation of the stars comprising the low-\afe{} sequence. The `radial migration' model of \citet{2009MNRAS.396..203S} assumes a continuous, smoothly-varying delivery of gas to the disc, but allows for the exchange of star-forming gas and stars between adjacent radial bins. This allows for the present-day co-location of disc stars whose formation conditions (and hence their location in \afe{}-\feh{} space) differed significantly. Bimodality of \afe{} at fixed \feh{} therefore stems from such a superposition of populations, fostered by the outward migration of high-\afe{} stars that formed rapidly close to the Galactic centre, and the inward migration of low-\afe{} stars formed farther out in the Galactic disc. 

Recently,  \citet{2016arXiv160408613A} used their flexible analytic chemical evolution model \textsc{flexCE}, to scrutinise the viability of these scenarios and assess their sensitivity to the variation of free parameters. They concluded that both models are able to produce high- and low-\afe{} sequences, but identified potential shortcomings in both: the two-infall model requires fine-tuning of the duration of the hiatus and may be incapable of yielding the bimodality in \afe{} over an extended a range of \feh{} seen by APOGEE \citep[see, e.g.][]{2014ApJ...796...38N,2015ApJ...808..132H}. Their implementation of the radial migration scenario produces a weaker bimodality than that revealed by APOGEE.

Analytic galaxy chemical evolution models such as those described above are an instructive means of assessing the impact of physical processes on the element abundance evolution of stellar populations. However, the scope and predictive power of such models is limited by their recourse to restrictive simplifications and approximations, such as the adoption of arbitrary inflow rates, simplistic gas distributions, and the assumption of complete and instantaneous mixing. Hydrodynamical simulations starting from cosmological initial conditions offer a complementary means of studying the origin of features such as the Galaxy's \afe{}-\feh{} distribution, since they are unencumbered by the most restrictive of these simplifications. The chief drawback of this approach has been the lack, to date, of realistic simulations capable of reproducing the key physical properties of the galaxy population, including the element abundance patterns of their disc stars.

Simulations of individual galaxies have successfully produced galaxies with old, thick stellar components with disc-like kinematics and, in cases where elemental abundances were tracked, enhanced $\alpha$ abundances \citep[e.g][]{2004ApJ...612..894B,2012MNRAS.426..690B,2013MNRAS.436..625S,2014MNRAS.442.2474M,2014MNRAS.443.2452M}. Examination of the ages of disc stars (for which \afe{} is often considered a proxy) in simulations has proven instructive, suggesting that the old, thick discs of Milky Way-like galaxies appear to form `upside-down' and `inside-out', such that discs are born thick, and gradually become thinner \citep[e.g][]{2006ApJ...639..126B,2013ApJ...773...43B,2016arXiv160804133M,2017arXiv170901040N}, likely in response to the decline of the gas accretion rate onto the galaxy and the concomitant decline of energy injection into the interstellar medium (ISM) from feedback. It has also been posited that `upside down' formation of a thick disc component may instead be a consequence of mergers between gas-rich clumps at early epochs \citep[e.g.][]{1998Natur.392..253N,2004ApJ...612..894B,2009ApJ...707L...1B}. Models in which elemental abundances are `painted' onto particles in N-body simulations successfully reproduce the geometry of mono-age and mono-abundance populations,  
\citep[e.g.][]{2013A&A...558A...9M,2015ApJ...804L...9M,2017ApJ...834...27M}.  Such models provide a means of understanding the presence of age gradients in the \emph{geometric} thick disc of the Milky Way \citep[e.g.][]{2016arXiv160901168M,2017arXiv170600018M}, as well as the appearance of radially extended thick disc components in external galaxies \citep[e.g.][]{2006AJ....131..226Y}. 

\citet{2017arXiv170807834G} recently reported the formation of disc star populations exhibiting both high- and low-\afe{} components in several hydrodynamical simulations of individual $\sim L^\star$ galaxies. Those authors studied 6 high-resolution `zoom' simulations from the Auriga suite \citep{2017MNRAS.467..179G}, and concluded that bimodal populations can form via two pathways. Bimodality in the inner disc is the result of a two-phase star formation history (SFH), characterized by a short but intense episode of star formation in the first Gyr or so, followed by a more prolonged and gentle SFH. In the outer disc, bimodality results from a brief cessation of star formation, associated with a phase of contraction of the early $\alpha$-rich gas disc, followed by star formation reignition.  The latter may be caused by accretion of fresh gas, predominantly associated with the merger of gas-rich satellites. Those authors remark that the formation of bimodal \afe{} sequences is not a universal outcome of their galaxy formation simulations, raising the questions of which haloes it arises in, and in response to which physical processes.

In this paper we present an analysis of the distribution on the \afe{}-\feh{} plane of the disc stars of Milky Way-like galaxies in the EAGLE cosmological simulations of galaxy formation \citep{2015MNRAS.446..521S,2015MNRAS.450.1937C}. The galaxy population formed by EAGLE has been shown to reproduce a broad range of observed galaxy properties and scaling relations, both at the present-day and at early cosmic epochs, such as the colour-magnitude relation \citep{2015MNRAS.452.2879T,2016MNRAS.460.3925T,2017MNRAS.470..771T}, the Tully-Fisher relation \citep{2017MNRAS.464.4736F} , and the evolution of galaxy sizes \citep{2017MNRAS.465..722F}. EAGLE has also been shown to reproduce the observed $\alpha$-enhancement of massive galaxies \citep{2016MNRAS.461L.102S}. The largest-volume EAGLE simulation follows the evolution of a periodic, cubic cosmic volume of $L=100\,{\rm cMpc}$ on a side, yielding a large population of galaxies, with diverse formation histories and present-day environments. Our objective is to examine whether galaxies with bimodal \afe{} at fixed \feh{} form within EAGLE and, if so, to establish the physical drivers underpinning their emergence. In the process, we exploit the statistics afforded by the EAGLE simulations to assess whether one should expect that the distribution of elemental abundance patterns exhibited by the Galactic disc is common amongst late-type galaxies of a similar mass, thus enabling us to interpret the findings of surveys such as APOGEE in the broader context of galaxy formation theory. 

The paper is organised as follows. In Section \ref{sec:methods}, we briefly summarise the EAGLE simulations and present our numerical methods. In Section \ref{sec:z0_props} we explore the \afe{}-\feh{} distribution of present-day Milky Way-like galaxies in EAGLE, examining correlations with the birth properties of stellar populations, the diversity of the distributions, and the frequency with which galaxies exhibit distinct \afe{}-\feh{} sequences. In Section \ref{sec:afeorigin}, we explore the origin of \afe{} bimodality by studying the gas infall and enrichment histories of EAGLE galaxies. In Section~\ref{sec:halo_accretion} we examine the connection between bimodal \afe{} distributions in galaxy discs, and the accretion history of their host dark matter haloes. We summarise our findings and discuss their broader implications in Section \ref{sec:summary_and_discussion}. Throughout, we adopt the convention of prefixing units of length with `c' and `p' to denote, respectively, comoving and proper scales, e.g. cMpc for comoving megaparsecs. 

\section{Numerical Simulations \& methods}
\label{sec:methods}

This section provides a brief overview of the simulations (Section \ref{sec:eagle}) and their subgrid physics routines (Section \ref{sec:subgrid_physics}). We focus in particular on the aspects of the implemented physics most relevant for this study, and direct the reader to the reference and methods papers of the EAGLE simulations for a more comprehensive description. Section \ref{sec:finding_galaxies} describes our methods for identifying and characterising galaxies. 

\subsection{The EAGLE simulations}
\label{sec:eagle}

The simulations analysed here are drawn from the EAGLE suite of cosmological, hydrodynamical simulations \citep{2015MNRAS.446..521S,2015MNRAS.450.1937C}, which model the formation and evolution of galaxies in a $\mathrm{\Lambda CDM}$ cosmogony described by the parameters advocated by the \citet{2014A&A...571A...1P}, namely {$\Omega_0 = 0.307$, $\Omega_{\rm b} =
0.04825$, $\Omega_\Lambda= 0.693$, $\sigma_8 = 0.8288$, $n_{\rm s} = 0.9611$, $h = 0.6777$, $Y = 0.248$}. The simulations were performed using a modified version of the smoothed particle hydrodynamics (SPH) and TreePM gravity solver \textsc{Gadget 3}, most recently described by \citet{2005MNRAS.364.1105S}. Modifications include the implementation of the pressure-entropy formulation of SPH presented by \citet{2013MNRAS.428.2840H}, the time-step limiter of \citet{2012MNRAS.419..465D}, and switches for artificial viscosity and artificial conduction of the forms proposed by, respectively, \citet{2010MNRAS.408..669C} and \citet{2010MNRAS.401.1475P}.

We examine several simulations from the EAGLE suite, primarily the simulation with the largest volume, Ref-L100N1504, which adopts the `Reference' model parameters \citep[see][]{2015MNRAS.446..521S} and follows a periodic cube of side $L = 100\,{\rm cMpc}$ with $1504^3$ collisionless dark matter particles of mass $9.70\times 10^6\,{\rm M}_\odot$ and an (initially) equal number of SPH particles of mass $1.81\times 10^6\,{\rm M}_\odot$. The simulations conducted at this `intermediate' resolution adopt a Plummer-equivalent gravitational softening length of $\epsilon_{\rm com} = 2.66\,{\rm ckpc}$, limited to a maximum proper length of $\epsilon_{\rm prop} = 0.7\,{\rm pkpc}$. To explore the enrichment history of galaxies at high temporal resolution, we also examine a realisation of Ref-L025N0376, (which has the same resolution as Ref-L100N1504, but follows a smaller $L=25\ \mathrm{cMpc}$ volume), for which 1000 full snapshots were recorded, rather than the usual 28, at an approximate spacing of 12 Myr. 

\subsubsection{Subgrid physics}
\label{sec:subgrid_physics}

The simulations adopt a metallicity-dependent density threshold for star formation \citep{2004ApJ...609..667S}. Gas particles denser than this threshold are eligible for stochastic conversion into stellar particles, with a probability that is dependent on their pressure \citep{2008MNRAS.383.1210S}. Supermassive black holes (BHs) are seeded in haloes identified by a friends-of-friends (FoF) algorithm run periodically during the simulation, and grow by gas accretion and mergers with other black holes \citep[see, e.g.][]{2005MNRAS.361..776S,2009MNRAS.398...53B,2015MNRAS.446..521S}. The rate of gas accretion onto BHs is influenced by the angular momentum of gas close to the BH \citep[see ][]{2015MNRAS.454.1038R} and does not exceed the Eddington limit.

Feedback associated with the evolution of massive stars (`stellar feedback') and the growth of BHs (`AGN feedback') is implemented as stochastic heating following \citet{2012MNRAS.426..140D}. Outflows develop without the need to specify an initial mass loading or velocity, and do not require that radiative cooling or hydrodynamic forces are temporarily disabled. The efficiency of stellar feedback is dependent upon the local density and metallicity of each newly-formed stellar particle to account, respectively, for residual spurious resolution-dependent radiative losses, and increased thermal losses in metal-rich gas. The dependence on these properties was calibrated to ensure that the simulations reproduce the present-day galaxy stellar mass function, whilst also yielding disc galaxies with realistic sizes \citep{2015MNRAS.450.1937C}. The efficiency of AGN feedback was calibrated to ensure that the simulations reproduce the present day scaling between the stellar masses of galaxies and the mass of their central BH. 

The mass of stellar particles is $\sim 10^6\,{\rm M}_\odot$, so each represents a population of stars and can be considered as a simple stellar population (SSP). We assume the initial distribution of stellar masses to be described by the \citet{2003PASP..115..763C} initial mass function (IMF) in the range $0.1-100\,{\rm M}_\odot$. The return of mass and nucleosynthesised metals from stars to interstellar gas is implemented as per \citet{2009MNRAS.399..574W}. The scheme follows the abundances of the 11 elements most important for radiative cooling and photoheating (H, He, C, N, O, Ne, Mg, Si, S, Ca and Fe), using nucleosynthetic yields for massive stars, Type Ia SNe, Type II SNe and the AGB phase from \cite{1998A&A...334..505P} and \cite{2001A&A...370..194M}. We use the metallicity-dependent stellar lifetimes advocated by \citet{1998A&A...334..505P}. The `lifetimes' of Type Ia SNe are described by an empirically-motivated exponential delay time distribution, such that their rate per unit initial stellar mass is:
\begin{equation}
\dot{N}_{\rm SNIa}(t) = \nu \frac{e^{-t/\tau}}{\tau},
\end{equation}
where $\nu = 2\times 10^{-3}\,{\rm M}_{\odot}^{-1}$ is the total number of Type Ia SNe per unit initial mass, and $\tau = 2\,{\rm Gyr}$ is the e-folding timescale. These parameters were calibrated to ensure that the simulations broadly reproduce the observed evolution of the cosmic Type Ia SNe rate density \citep{2015MNRAS.446..521S}.  

At each timestep, the mass and metals released from evolving stellar populations are transferred from stellar particles to their SPH neighbours according to the SPH kernel (for which we use the $C^2$ kernel of Wendland 1995), with weights calculated using the initial, rather than current, mass of the particle \citep[see Section 4.4 of][]{2015MNRAS.446..521S}. The transferred mass is `fixed' to SPH particles and does not diffuse. To alleviate the symptoms of this suppressed mixing, gas particles also carry a kernel-smoothed measurement of each element abundance, which is updated at each active timestep \citep[for a detailed discussion, see][]{2009MNRAS.399..574W}. Following the implementation of \citet{2009MNRAS.393...99W}, smoothed abundances are used to compute, element-by-element, the rates of radiative cooling and heating of gas in the presence of the cosmic microwave background and the metagalactic UV background due to the galaxies and quasars, as modelled by \citet{2001cghr.confE..64H}. For the purposes of this calculation, the gas is assumed to be optically thin and in ionisation equilibrium. Stellar particles inherit the elemental abundances of their parent gas particle. Throughout, we present measurements using the smoothed abundances mentioned above. Despite the absence of element diffusion between particles, the mixing of particles with differing abundances is a form of diffusion that is modelled by our simulations, accessed via the use of smoothed abundances.

The simulations do not explicitly model the cold, dense phase of the ISM, and thus impose a temperature floor, $T_{\rm eos}(\rho)$, to prevent spurious fragmentation within star-forming gas. The floor takes the form of an equation of state $P_{\rm eos} \propto \rho^{4/3}$ normalised so $T_{\rm eos} = 8000 {\rm K}$ at $n_{\rm H} = 0.1 {\rm cm}^{-3}$. The temperature of star-forming gas therefore reflects the effective pressure of the ISM, rather than its actual temperature. Since the Jeans length of gas on the temperature floor is $\sim 1\,{\rm pkpc}$, a drawback of its use is that it suppresses the formation of gaseous discs with vertical scale heights much shorter than this scale; moreover, as recently shown by \citet{2018MNRAS.473.1019B}, self-gravitating discs in EAGLE are likely vertically thickened since the gravitational softening length is similar to the disc scale height. We comment further on the consequences of this limitation in Section \ref{sec:disc_stellar_pops}. 

\subsection{Identification and characterisation of galaxies}
\label{sec:finding_galaxies}
Galaxies and their host haloes are defined by a two-step process. Haloes are identified by applying the FoF algorithm to the dark matter particle distribution, with a linking length of 0.2 times the mean interparticle separation. Gas, stars and BHs are assigned to the FoF group, if any, of their nearest dark matter particle. Bound substructure within haloes, comprised of any particle type, is then identified using the \textsc{Subfind} algorithm \citep{2001MNRAS.328..726S,2009MNRAS.399..497D}. For each FoF halo, the subhalo comprising the most-bound particle is defined as the central subhalo, all other subhaloes are defined as satellites.

In general, unless stated otherwise, the properties of the `galaxy' associated to a given subhalo are defined by aggregating the properties of the particles that are bound to the subhalo and also reside within a spherical aperture of radius $r = 30 {\rm pkpc}$, centred on the subhalo's most-bound particle. For galaxies of the mass we examine here, this aperture mimics the 2-dimensional Petrosian aperture widely used in observational studies \citep[see][]{2015MNRAS.446..521S}.

To characterise the morphology of EAGLE galaxies, we follow \citet{2017arXiv170406283C} and compute the fraction of the kinetic energy of a galaxy's stellar particles invested in ordered co-rotation with the disc:
\begin{equation}
\kappa_{\mathrm{co}} = \frac{K_{\mathrm{rot}}}{K} = \frac{1}{K}\sum^{r < 30\mathrm{pkpc}}_i{\frac{1}{2}m_i\left(\frac{L_{Z,i}}{m_{i}R_{i}}\right)^{2}},
\end{equation}
where $m_i$ is the mass of the $i^{\mathrm{th}}$ particle, $L_Z$ is the z-component of its angular momentum , and $R$ is the cylindrical radius in the disc plane of the particle position with respect to the galaxy centre. \citet{2017arXiv170406283C} show that a threshold of $\kappa_{\mathrm{co}} = 0.4$ broadly separates morphologically disc-like galaxies with blue intrinsic $u$-$r$ colour, from redder, more elliptical galaxies. 

Following \citet{2016MNRAS.461L.102S}, we use $\mathrm{[O/Fe]}$ as a proxy for \afe{}, since oxygen dominates the mass budget of $\alpha$ elements. We use the common definition of abundance ratios $\mathrm{[x/y]}$ relative to solar values,
 \begin{equation}
 \left[\frac{\mathrm{x}}{\mathrm{y}}\right]  = \log_{10}{\left(\frac{X^{\mathrm{x}}}{X^{\mathrm{y}}}\right)} -  \log_{10}{\left(\frac{X_{\odot}^{\mathrm{x}}}{X_{\odot}^{\mathrm{y}}}\right)},
 \end{equation}
where $(x,y)$ each represent an element and $X^{\mathrm{x}} = $ denotes the galaxy stellar mass fraction comprised by element x. We adopt the solar mass fractions of \citet{2009ARA&A..47..481A}, who report $X_{\odot}^{\mathrm{O}}/X_{\odot}^{\mathrm{Fe}} = 4.76$ and $X_{\odot}^{\mathrm{Fe}}/X_{\odot}^{\mathrm{H}} = 0.0011$.

\subsubsection{Defining the stellar populations of galaxy discs}
\label{sec:disc_stellar_pops}

To facilitate a like-for-like comparison of elemental abundances inferred from Galactic surveys with those of simulated galaxies broadly similar to the Milky Way, we construct samples of galaxies with present day stellar mass in the interval $M_\star = (5 - 7) \times 10^{10} {\rm M}_\odot$, broadly similar to the value of $\simeq 6 \times 10^{10}~{\rm M}_\odot$ estimated for the Galaxy \citep[e.g.][]{2011MNRAS.414.2446M,2016ARA&A..54..529B}, that are also disc-dominated ($\kappa_{\mathrm co} > 0.4$). These criteria are satisfied by 133 galaxies in Ref-L100N1504, and 5 in Ref-L025N0376. The mean stellar half mass radius of galaxies comprising this `Milky Way-like' sample in Ref-L100N1504 is $R_{1/2} \simeq 7.5 {\rm pkpc}$, consistent with the average $r$-band scale length of external disc galaxies in SDSS, $R_{1/2} = 5.7 \pm 1.9$ kpc, reported by \citet{2010MNRAS.406.1595F}.

We mimic, crudely, the selection function of Galactic surveys such as Gaia-ESO and APOGEE by considering only those stellar particles within a cylindrical annulus, centred on the most-bound particle of the galaxy, with inner and outer radii equal to half and twice the stellar half mass radius (of the individual galaxy), respectively, and upper and lower vertical bounds of $\pm$ one quarter of the stellar half mass radius. This selects roughly $10^4$ particles per galaxy. Throughout, unless otherwise stated, we define `disc stars' as the stellar particles bound to `Milky Way-like' galaxies, satisfying this geometric constraint. No kinematic constraints are applied to the stellar particles. We do not dissect the \afe{}-\feh{} distribution into radial and vertical bins since, for reasons articulated in Section \ref{sec:subgrid_physics}, the scale heights of the young and old disc stellar populations are necessarily more similar in the simulations than is observed by Galactic surveys. 

\section{The elemental abundances of disc stars in Milky Way-like galaxies}
\label{sec:z0_props}

In this section, we examine the distribution, in the \afe{}-\feh{} plane, of the disc stars of Milky Way-like galaxies, and explore the relationship between this distribution and the underlying properties of the stellar population. We refrain from performing a detailed comparison of the \afe{}-\feh{} distribution with those recovered from surveys of the Galaxy's stellar populations, since our aim is to understand the origin of the trends in the distribution, rather than to reproduce the observed distribution precisely. Predictions from models, and inferences from observations, of absolute (as opposed to relative) abundances are subject to systematic uncertainties of a factor $\gtrsim 2$. The uncertainties stem primarily from theoretical uncertainties in nucleosynthetic yield calculations \citep[see e.g. Appendix A of][and references therein]{2009MNRAS.399..574W}, the calibration of observational abundance indicators \citep[e.g.][]{2008ApJ...681.1183K}, statistical and systematic uncertainties in the measurement of the volumetric Type Ia SNe rate \citep[e.g.][]{2008ApJ...681..462D,2010ApJ...713.1026D,2014ApJ...783...28G}, and an incomplete understanding of the nature of Type Ia SNe progenitors and their delay-time distribution \citep[see, e.g.][]{2012NewAR..56..122W}. In Appendix \ref{sec:subgrid}, we show the effect on the \afe{}-\feh{} distribution of varying the subgrid parameters governing the number of Type Ia SNe per unit stellar mass formed, and their e-folding timescale. There, we show that the distribution can change significantly in response to reasonable variation of these parameters.

\subsection{The distribution of disc stars on the \afe{}-\feh{} plane}
\label{sec:stacked_afe_feh}

\begin{figure}
\includegraphics[width=\columnwidth]{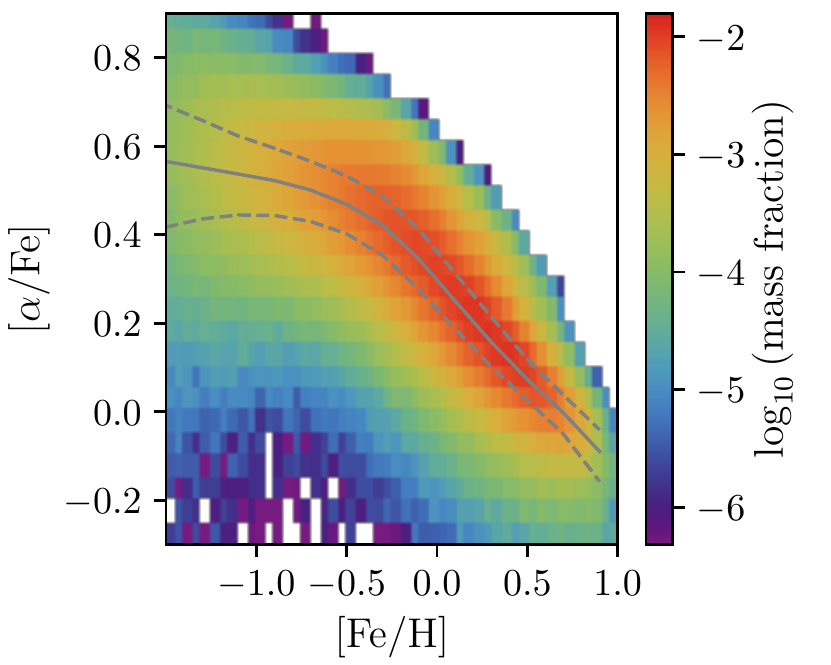}
\caption{\label{fig:afestack} Two-dimensional histogram of the mass-weighted \afe{}-\feh{} distribution of all `disc stars' associated with the 133 galaxies identified as broad analogues of the Milky Way in terms of their stellar mass and morphology (see Section \ref{sec:disc_stellar_pops}) at $z=0$ in Ref-L100N1504. The overplotted solid line shows the median \afe{} in bins of $\Delta$\feh{}\,$=0.2$, dashed lines show the interquartile range.}
\end{figure}

Fig. \ref{fig:afestack} shows the \afe{}-\feh{} distribution of the 1.5 million stellar particles comprising the disc populations (see Section \ref{sec:disc_stellar_pops}) of the 133 present-day Milky Way-like galaxies in Ref-L100N1504 as a 2-dimensional histogram. The pixels of the histogram represent bins of 0.05 in $\Delta$\afe{} and $\Delta$\feh{}, and the value of each pixel is weighted by the current (rather than initial) mass of the stellar particles within. The overplotted filled line represents the median \afe{} calculated in bins of $\Delta$\feh{}$=0.2$, and the dashed lines show the interquartile range. As has been observed by Galactic surveys, and is commonly predicted by analytic and numerical Galactic chemical evolution models, the primary trend is that of a sequence with declining \afe{} as a function of \feh{}, with a relatively shallow negative gradient at low \feh{} and steeper gradient at higher \feh{}. The distribution of \afe{} at fixed \feh{} is unimodal, with a broad dispersion. The median \afe{} declines gradually from $0.57$ at \feh{}$\,=-1.0$ to $\simeq 0.47$ at \feh{}$\,=-0.5$, then declines more rapidly to \afe{}$\,\simeq 0.07$ at \feh{}$\,=0.5$. The $1\sigma$ scatter in \afe{} is approximately $0.41$ at \feh{}$\,=-1.0$, and narrows to $0.14$ at \feh{}$\,=0.5$. The increased scatter at low \feh{} is likely a consequence of poor sampling of the enrichment process.

\begin{figure}
\includegraphics[width=\columnwidth]{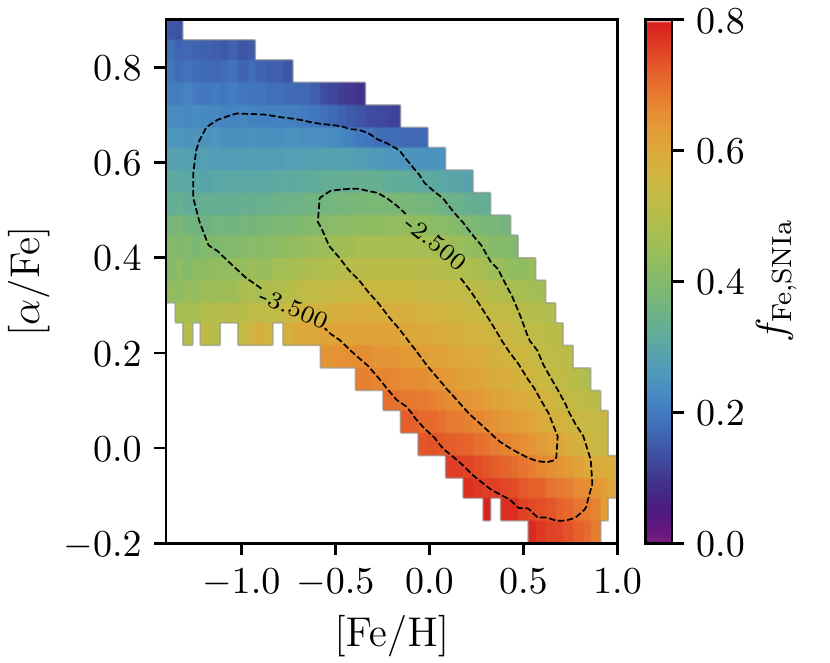}
\caption{\label{fig:afesniafrac} The mean mass fraction of the Fe, locked up in the disc stars of present-day Milky Way-like galaxies in Ref-L100N1504, that was synthesised by Type Ia SNe. The fraction is shown as a function of the stellar populations' position in \afe{}-\feh{} space. The value in each  pixel is weighted by the current mass of the stellar particles within. Overplotted contours reproduce the mass distribution shown in Fig. \ref{fig:afestack}. The Type Ia SNe Fe fraction broadly anti-correlates with \afe{}, and at fixed \afe{} the Fe mass fraction contributed by Type Ia SNe is greatest in Fe-poor stars.}
\end{figure}

We next turn to an examination of the underlying properties of the disc stars, to elucidate the origin of the distribution shown in Fig. \ref{fig:afestack}. The common interpretation for the diversity of \afe{} in the Galaxy's disc stars is a varying relative contribution of Type Ia and Type II SNe ejecta to the Fe abundance of each star. This fraction is tracked explicitly by the simulations, enabling this hypothesis to be tested directly. We examine the same sample of stellar particles shown in Fig. \ref{fig:afestack}, and show in Fig. \ref{fig:afesniafrac} the mean mass fraction of their Fe that was synthesised by Type Ia SNe, $f_{\rm Fe,SNIa}$, as a function of their position in \afe{}-\feh{} space. The mean fraction of each pixel is computed weighting by the current mass of its contributing stellar particles. The mass distribution shown in Fig. \ref{fig:afestack} is illustrated here with overlaid contours, the outer and inner contours corresponding to $\log_{10}{(\mathrm{mass\ fraction})} = -3.5$ and $-2.5\ \mathrm{pixel^{-1}}$, respectively. Only well sampled pixels, with $\log_{10}{(\mathrm{mass\ fraction})} > -4.5\ \mathrm{pixel^{-1}}$, corresponding to approximately 50 stellar particles, are shown.

As one might expect, there is a broad anti-correlation between \afe{} and $f_{\rm Fe,SNIa}$. According to our simulations the majority of the Fe locked into disc stars with \afe{}$\,\gtrsim 0.5$ was synthesised by Type II SNe, with the mass fraction synthesised in Type Ia SNe being typically $< 0.2$. Conversely, in stars with subsolar \afe{}, which typically exhibit supersolar \feh{}, the mass fraction of Fe synthesised by Type Ia SNe can be $\simeq 0.8$. There is also a weaker, but significant trend at fixed \afe{} ($\lesssim 0.2$), such that stars with higher \feh{} have a smaller Fe contribution from Type Ia SNe. Enrichment of gas by Type II SNe increases \feh{} at broadly fixed \afe{} (a shift to the right in the \afe{}-\feh{} plane) if enrichment by Type Ia is negligible, or whilst increasing \afe{} (a shift up and to the right) if Type Ia enrichment is significant. Enrichment of Type Ia SNe tends to increase \feh{} whilst decreasing \afe{} (a shift down and to the right). Therefore, greater \feh{} at fixed \afe{} is typically due to the more Fe-rich stars sourcing a greater fraction of their Fe from from Type II SNe relative to Type Ia SNe. 

\begin{figure}
\includegraphics[width=\columnwidth]{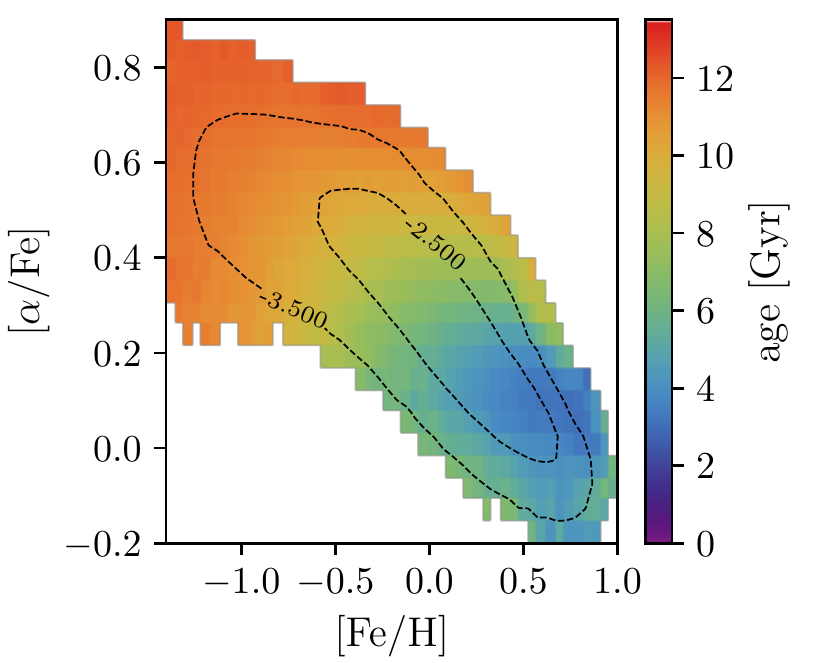}
\caption{\label{fig:afeages} The mean age of the disc stars of present day Milky Way-like galaxies in Ref-L100N1504, as a function of their position in \afe{}-\feh{} space. Pixel values are weighted by the current mass of the stellar particles within, and the overplotted contours reproduce the mass distribution shown in Fig. \ref{fig:afestack}. Age correlates with \afe{}, though at any fixed \afe{} disc stars exhibit a broad range of mean ages, depending on their \feh{}. At fixed \feh{}, the most $\alpha$-rich stellar populations tend to be the oldest.}
\end{figure}

\begin{figure}
\includegraphics[width=\columnwidth]{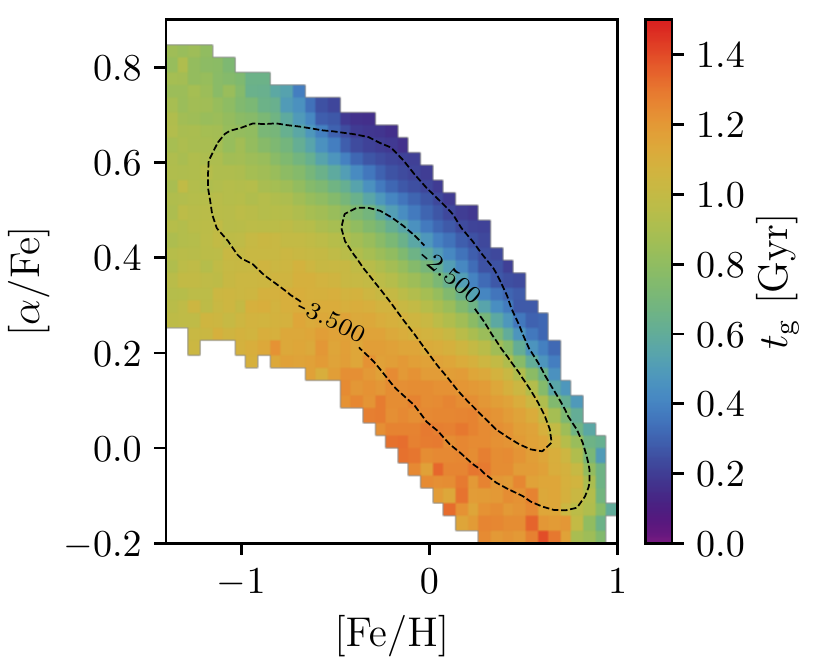}
\caption{\label{fig:afetcon} The mean gas consumption timescale $t_{\mathrm{g}}$ of the natal gas from which the disc stars of present-day Milky Way-like galaxies in Ref-L100N1504 formed, as a function of their position in \afe{}-\feh{} space.  Pixel values are weighted by the current mass of the stellar particles within, and the overplotted contours reproduce the mass distribution shown in Fig. \ref{fig:afestack}. At fixed \feh{}, the most $\alpha$-rich stellar particles formed from gas with the shortest consumption timescales.}
\end{figure}

Analytic models posit that high-\afe{} stellar populations form in the early life of a galaxy, and do so rapidly, such that there is little opportunity for the enrichment of star-forming gas by the delayed release of Type Ia SNe ejecta. \citet{2016MNRAS.461L.102S} show that this is the case for the galaxy-averaged $\alpha$-enhancement of massive EAGLE galaxies, whose stars form rapidly at early times prior to quenching by AGN feedback. To examine whether the same applies in Milky Way-like galaxies, we plot in Fig. \ref{fig:afeages} the mean age of stellar particles as a function of their position in \afe{}-\feh{} space, and in Fig. \ref{fig:afetcon} the mean consumption timescale\footnote{The inverse of the consumption timescale is often referred to as the `star formation efficiency' in the chemical evolution modelling literature.} of the natal gas, $t_{\rm g} = \Sigma_{\rm g}/\dot{\Sigma}_\star$, where $\Sigma_{\rm g}$ is the gas surface density and $\dot{\Sigma}_\star$ is the star formation rate (SFR) per unit area. As in previous plots, the pixel values are weighted by the current mass of the stellar particles within, and overlaid contours represent the mass distribution shown in Fig. \ref{fig:afestack}. 

The consumption timescale provides an estimate of the period of time a parcel of star-forming gas resides in the ISM, during which it can be enriched by the ejecta of neighboring stellar populations. It is likely that $t_{\rm g}$ overestimates the time spent in the ISM by a star-forming particle, since it only considers the regulation of the gas surface density by star formation, thus neglecting the influence of feedback. On average, $\langle N_{\rm heat} \rangle \simeq 1$ SPH particle is stochastically heated by the stellar feedback accompanying the formation of each stellar particle, and these particles can entrain neighboring particles in outflows. Outflows can also be driven by AGN feedback, so the accuracy of this estimate is likely poorer than a factor of $\simeq 2$. Nonetheless, $t_{\rm g}$ remains an instructive diagnostic for our purposes. To compute it, we follow \citet{2008MNRAS.383.1210S} and assume that the scale height of star-forming gas discs is comparable to the local Jeans length, $L_{\rm J}$, such that $\Sigma_{\rm g} \simeq \rho_{\rm g}L_{\rm J}$, and then relate the consumption timescale of star-forming gas to its pressure:
\begin{equation}
\label{eq:t_g}
t_{\rm g} = A^{-1}(1\,{\rm M}_\odot\,{\rm pc}^{-2})^n\left( \frac{\gamma}{G} f_{\rm g}P_\star\right)^{(1-n)/2}.
\end{equation}
Here, $\gamma=5/3$ is the ratio of specific heats for an ideal gas, and $f_{\rm g}$ is the local gas fraction, which we assume to be unity. The parameters $A$ and $n$ are specified by observations, i.e. the Kennicutt-Schmidt scaling relation. We use $A=1.515\times 10^{-4}\,{\rm M}_\odot\,{\rm yr}^{-1}\,{\rm kpc^{-2}}$ and $n=1.4$, with the former being a factor of $1.65$ lower than the value specified by \citet{1998ApJ...498..541K}, since we assume a Chabrier, rather than Salpeter, IMF. For the purposes of the calculation, the pressure of the natal gas, $P_\star$, is assumed to be that specified by the temperature floor described in Section \ref{sec:eagle}, i.e. $P_\star = P_{\rm eos}(\rho_\star)$, where $\rho_\star$ is the density of the natal gas at the instant it is converted into a stellar particle.

Fig. \ref{fig:afeages} shows that, as expected, there is a strong correlation between the present-day age of stellar populations, and their position in \afe{}-\feh{} space. The characteristic age of Fe-poor (\feh{}$\,\lesssim -0.5$), $\alpha$-rich (\afe{}$\,\gtrsim 0.4$) stars is greater than $10\,{\rm Gyr}$, corresponding to a formation redshift of $z \gtrsim 1.7$, whilst those with solar or supersolar iron abundance, for which \afe{}$\,\lesssim 0.2$, are typically younger than $5\,{\rm Gyr}$ ($z_{\rm form} \lesssim 0.5$). There is a clear preference for $\alpha$-rich stars to be old, but disc stars can exhibit a broad range of ages at fixed \afe{}, such that there is not a direct mapping between age and $f_{\rm Fe,SNIa}$. This notwithstanding, at any fixed value of \feh{}, the stars richest in $\alpha$ elements generally tend to be the oldest. 

Fig. \ref{fig:afetcon} reveals a striking trend, clearly highlighting that at fixed \feh{} the most $\alpha$-rich populations form from gas that exhibits the shortest consumption timescales. As the dynamic range is narrow for $t_{\rm g} \gtrsim 1\,{\rm Gyr}$, we adopt a linear scaling of colour with $t_{\rm g}$. The plot illustrates clearly that the reason the release of ejecta from Type Ia SNe is able to influence the \afe{} of stellar populations at fixed \feh{} is the fact that the gas consumption timescales are of the same order as the characteristic e-folding timescale of the Type Ia SNe delay time distribution. Although gas can be enriched at densities below the star formation density threshold (whence $t_{\rm g}$ is infinite), the clear connection between $t_{\rm g}$ and (in particular) \afe{} highlights that, for most stellar particles, the bulk of their enrichment must take place whilst they comprise the star-forming ISM. The e-folding timescale of $\tau=2\,{\rm Gyr}$ adopted by the Reference model corresponds to a `halflife' of $t_{1/2}=\tau\ln(2)=1.4\,{\rm Gyr}$. The adoption of a shorter (longer) e-folding timescale results in the advancement (delay) of the release of Fe synthesised by Type Ia SNe, thus inhibiting (promoting) the formation of disc stars with high \afe{}. 

The majority ($\gtrsim 85$ percent) of stellar particles formed from gas with consumption times $\gtrsim 1\,{\rm Gyr}$, but those that formed more rapidly were largely precluded from enrichment by the Type Ia SNe of recently-formed, nearby stellar populations. We note that the $t_{\rm g}$ distribution on the \afe{}-\feh{} plane does not map directly onto that of the distribution of $f_{\rm Fe,SNIa}$ shown in Fig. \ref{fig:afesniafrac}. This is because short consumption timescales are realised within gas-rich overdensities at early epochs, and also within massive, metal-rich galaxies at later times.

\subsection{Galaxy-to-galaxy diversity of $\alpha$-enrichment}
\label{sec:afefeh_diversity}

Having examined the distribution of stars in \afe{}-\feh{} space, in a collective sense, for the present-day Milky Way analogues identified in Ref-L100N1504, we now examine the galaxy-to-galaxy diversity of the \afe{}-\feh{} distribution. This exercise enables a first assessment of how common among other present-day disc galaxies is the observed distribution in the \afe{}-\feh{} plane of the Milky Way's disc stars.

\begin{table}
\center
\caption{\label{tab:diag} Basic properties of the three present-day galaxies shown in Figure \ref{fig:afeexamples}. The rows correspond to, respectively: the labels applied to each galaxy in the text, their Galaxy ID and FOF ID in public EAGLE galaxy catalogues, their stellar mass, halo mass, their specific star formation rate (sSFR), their  $\kappa_{\mathrm{co}}$ value, and their half-mass radius.}
\begin{tabular}[]{r|rrrl}
\hline
\hline
Label                               & A    & B    & C    &  \\
Galaxy ID							& 16850421 & 16925427 & 16921468  & \\
FOF ID                              & 507  & 527  & 526  &  \\
$M_{*}$                             & 5.18 & 5.80 & 6.52 & $[10^{10} {\rm M}_{\odot}$] \\
$M_{200}$                           & 2.32 & 2.92 & 3.41 & $[10^{12} {\rm M}_{\odot}$] \\
sSFR                                & 4.74 & 5.14 & 1.85 & [$ 10^{-11} \mathrm{yr^{-1}}$] \\
$\kappa_{\mathrm{co}}$            & 0.67 & 0.45 & 0.46 & \\
$r_{1/2}$                           & 8.83 & 9.49 & 3.00 & [$\mathrm{kpc}$] \\
\hline
\end{tabular}

\end{table}

The present-day \afe{}-\feh{} distributions of the sample of 133 EAGLE galaxies exhibit significant diversity. Based on visual inspection, we have broadly classified the galaxies into the following categories, with the occupancy of each category in parentheses: unimodal with low-\afe{} (82), unimodal with high-\afe{} (19), broad \afe{} at fixed \feh{} (10), ambiguous (22). The sample is therefore dominated by galaxies that exhibit a single, broadly continuous `sequence' that is (unsurprisingly) similar to the overall trend revealed in Fig. \ref{fig:afestack}. A handful of single-sequence galaxies track the upper envelope of the stacked distribution and might reasonably be considered analogous to the stellar populations comprising the high-\afe{} sequence observed in the Galaxy. Six of the 10 galaxies categorised as having broad \afe{} at fixed \feh{} in fact exhibit a clear bimodality in \afe{} at fixed \feh{}, and might be considered, somewhat subjectively, as qualitatively similar to the Galaxy. The sample also includes systems with complex abundance distributions that in some cases are indicative of a recent merger with a gas-rich companion, for example where \afe{} exhibits a positive correlation with \feh{} over a narrow range.

\begin{figure*}
\includegraphics[width=1.\textwidth]{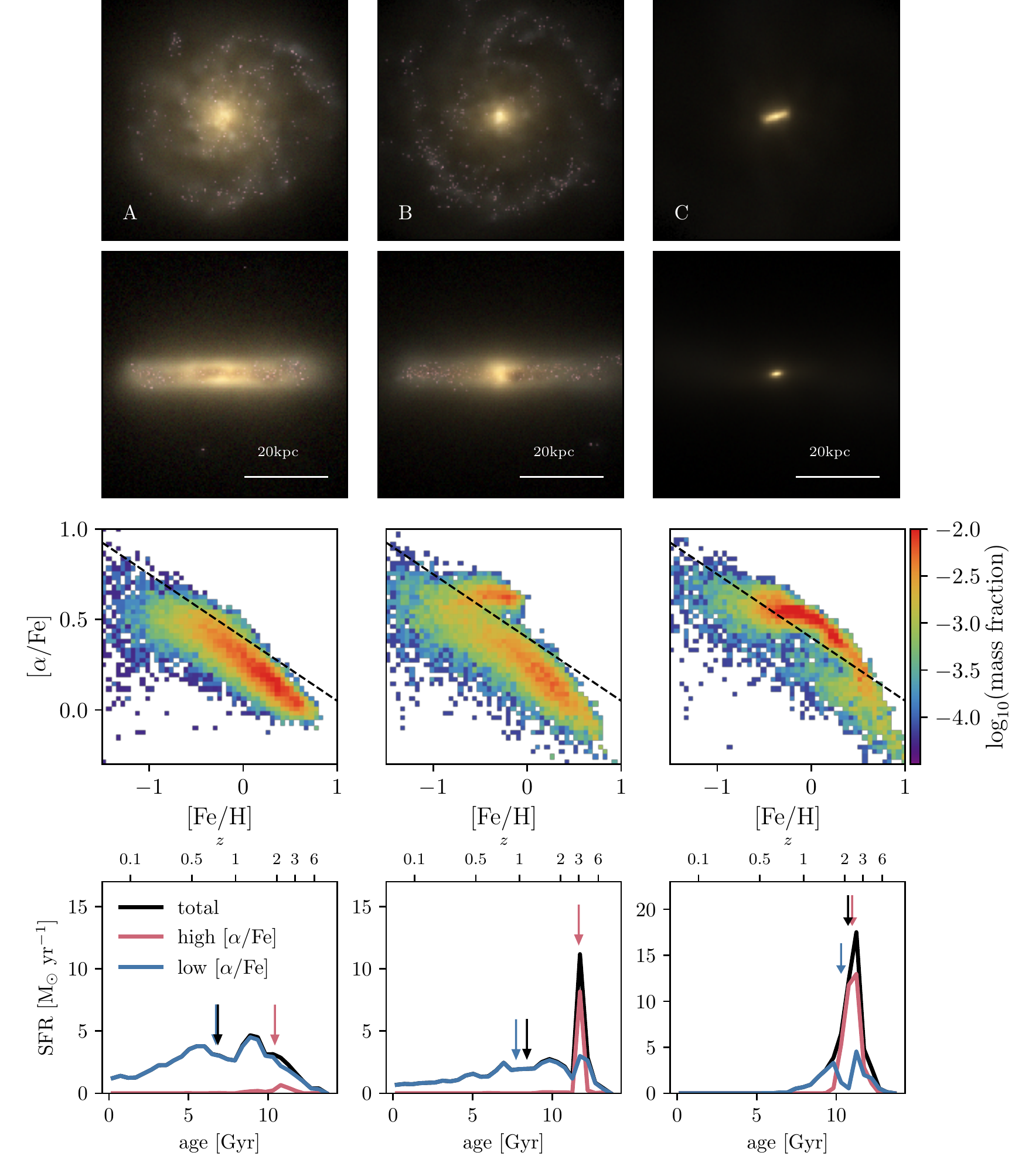}
\caption{\label{fig:afeexamples} Examples illustrating the diversity of the \afe{}-\feh{} distribution of disc stars from present-day galaxies in Ref-L100N1504. From left to right, the columns correspond to the galaxies labelled A, B and C in the text, for which key properties are quoted in Table \ref{tab:diag}. The two upper rows show mock images of the galaxies, with a $50 \times 50\,{\rm kpc}$ field of view, in the face-on and edge-on orientations. The images show the stellar light based on the combination of monochromatic $u$-, $g$- and $r$-band SDSS filters. The third row shows the \afe{}-\feh{} plane of the disc stars, with the dashed diagonal line positioned to broadly separate the high- and low-\afe{} sequences. This separation enables, in the bottom row, the contribution to the total (\textit{black}) star formation history of the stars comprising the high- (\textit{red}) and low-\afe{} (\textit{blue}) sequences to be shown. Downward arrows denote the epoch by which half of the stellar particles comprising these populations formed.}
\end{figure*}

To illustrate the diversity of the \afe{}-\feh{} distributions, we identify three representative examples: a galaxy exhibiting a single low-\afe{} sequence, a galaxy exhibiting bimodality in \afe{} at fixed \feh{}, and a galaxy exhibiting a single, high-\afe{} sequence. Key properties of these galaxies, labelled A, B and C, respectively, are given in Table \ref{tab:diag}. From top to bottom these are, respectively, the FOF halo identifier of the galaxy in the EAGLE public database\footnote{\url{http://galaxy-catalogue.dur.ac.uk}} \citep{2016A&C....15...72M}; its stellar mass, $M_\star$; its virial mass, $M_{200}$, defined as the total mass (in gas, stars, BHs and dark matter) enclosed by a sphere of radius $R_{200}$, centred on the galaxy's most bound particle, within which the mean enclosed density is $200$ times the critical density, $\rho_{\rm c} \equiv 3H^2/8\pi G$; the specific star formation rate of the galaxy, $\dot{M}_\star/M_\star$; its median kinetic energy in ordered rotation, $\langle \kappa \rangle$; and its stellar half-mass radius, $R_{1/2}$. 

The columns of Fig. \ref{fig:afeexamples}, from left to right respectively, show further properties of A, B and C. The two upper rows show images of the galaxies at $z=0$ with a $50 \times 50\,{\rm kpc}$ field of view, in face-on and edge-on orientation, defined such that the disc plane is orthogonal to the angular momentum axis of the stellar particles within $30\,{\rm kpc}$ of the most-bound particle. The images were extracted from the EAGLE public database, and were created using the techniques described by \citet{2015MNRAS.452.2879T}. The third row presents the \afe{}-\feh{} distribution of the disc stars, shown as a mass-weighted 2-dimensional histogram. The dashed diagonal line is an arbitrary threshold chosen to separate the high- and low-\afe{} sequences, which enables the contribution to the star formation history of the particles comprising each sequence to be shown separately in the bottom row. The downward arrows denote the epoch by which half of the stellar particles comprising the populations formed. 

Galaxies A and B exhibit a classical structure, with a central red spheroid surrounded by an extended blue disc. The spheroid of galaxy A is relatively diffuse, whilst that of galaxy B is more massive and more compact. Galaxy C is dominated by a central, rapidly-rotating elongated spheroid, and is an example of the relatively rare cases for which a high value of $\langle \kappa \rangle$ does not correspond to a morphologically-extended disc \citep[see][]{2017arXiv170406283C}. The bulk of galaxy A's disc stars form a continuous sequence in \afe{}-\feh{} space, at a relatively low value of \afe{}, extending to \feh{}$\,\simeq 0.8$. A similar sequence is visible in the case of galaxy B, extending almost to \feh{}$\,\simeq 1$. This low-\afe{} sequence is supplemented by an $\alpha$-rich (\afe{}$\,\simeq 0.65$) sequence that extends to  \feh{}$\,\simeq 0$. Galaxy C is dominated by a high-\afe{} sequence that is approximately constant at \afe{}$\,\simeq 0.6$ until \feh{}$\,\simeq 0$, and gradually declines as a function of increasing \feh{}. A small fraction of galaxy C's disc stars populate a region of \afe{}-\feh{} space similar to galaxy B's low-\afe{} sequence at the highest values of \feh{}.

The formation histories of the stars comprising the sequences offer clues to the latter's origin. Galaxy A exhibits an extended star formation history that evolves smoothly between values of $1-5\,{\rm M}_\odot\,{\rm yr}^{-1}$. This history is consistent with relatively long consumption timescales ($\gtrsim 1\,{\rm Gyr}$), and yields a relatively young stellar disc (an initial mass-weighted mean age of $7\ {\rm Gyr})$. The low-\afe{} component of galaxy B behaves similarly, albeit at a slightly lower SFR than galaxy A. The formation of galaxy B's high-$\alpha$ component dominates the early stages of its disc formation, and during this period of rapid star formation the SFR peaks at $\dot{M}_\star > 10\,{\rm M}_\odot {\rm yr}^{-1}$. The stars comprising this sequence therefore formed with an initial mass-weighed mean consumption time of $t_{\rm g} \simeq 200\ {\rm Myr}$. Nearly all of the stellar particles comprising the high-\afe{} sequence are formed during this episode. The formation of galaxy C is similar to the early behaviour of galaxy B, albeit more extreme. The massive, concentrated spheroid forms in a single, extended episode of rapid star formation during which the SFR peaks at $\dot{M}_\star \simeq 17\,{\rm M}_\odot {\rm yr}^{-1}$, yielding a initial mass-weighed mean consumption time of  $t_{\rm g} \simeq 180\ {\rm Myr}$. 

The star formation histories of these examples therefore corroborate the broad picture inferred from inspection of Figs. \ref{fig:afesniafrac}, \ref{fig:afeages} and \ref{fig:afetcon}: high values of \afe{} are realised by stellar populations with only a small fraction of their Fe mass synthesised by Type Ia SNe. Such populations are typically formed at early cosmic epochs, from gas with $t_{\rm g} < \tau$. This close connection between the star formation history and the distribution of stars in the \afe{}-\feh{} plane highlights that the former could be `reverse engineered' by applying analytic chemical evolution models to measurements of the latter, but the conclusions drawn from such an exercise might not be generally applicable to the broader population of galaxies with similar mass and morphology. 

\section{The origin of the \afe{}-\feh{} distribution of disc stars}
\label{sec:afeorigin}

In this section we examine the evolution of the gas from which the stellar populations occupying specific regions of the \afe{}-\feh{} plane formed. We also examine the subsequent radial migration of these populations within the disk. We begin by analysing a single galaxy, and generalise the analysis to the broader population in Section \ref{sec:broader_applicability}.

\begin{figure*}
\includegraphics[width=\textwidth]{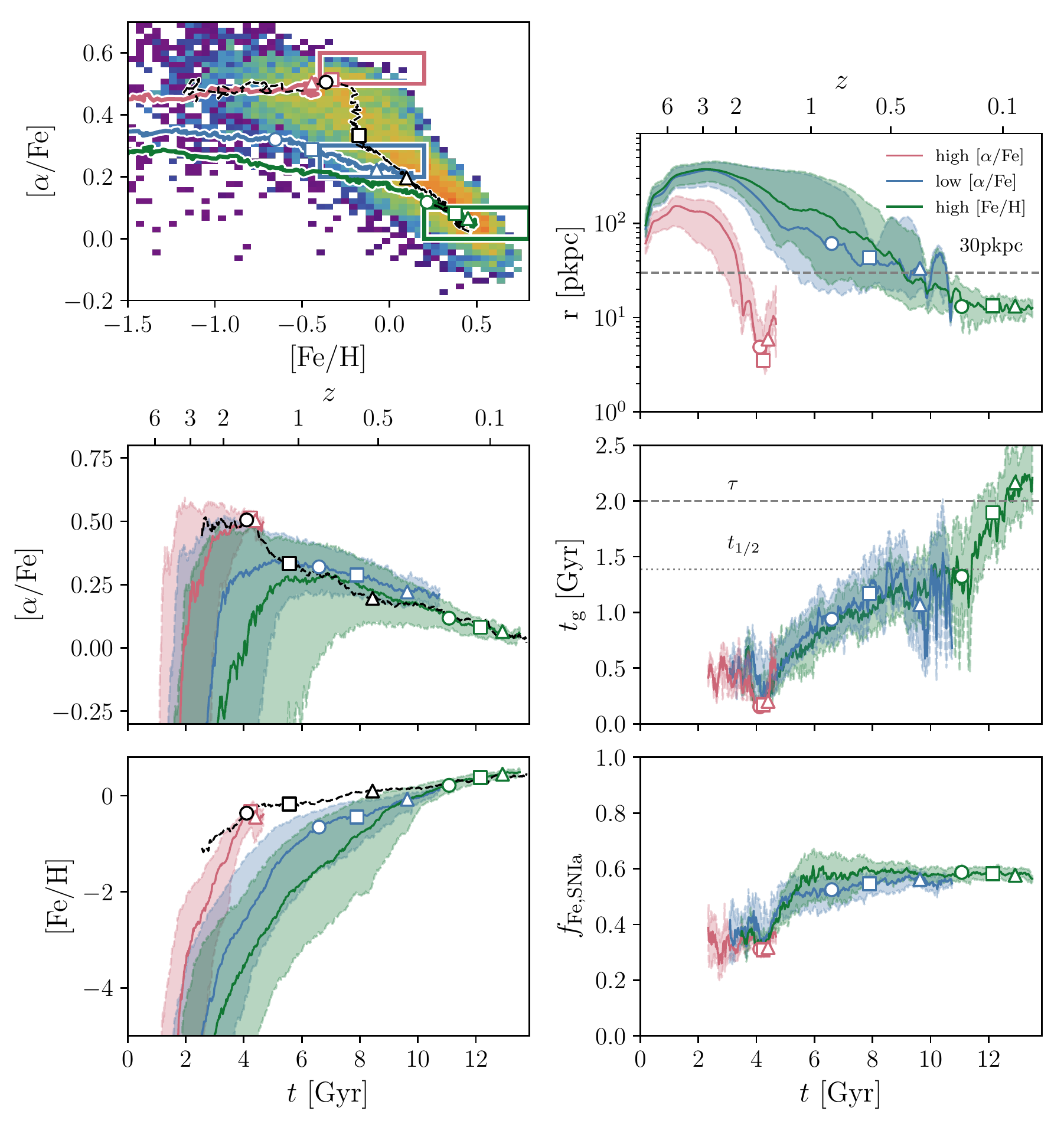}
\caption{\label{fig:daddyplot} The enrichment history of the natal gas of disc stars occupying selected regions of the $z=0$ \afe{}-\feh{} plane, for the galaxy discussed in Section \ref{sec:afeorigin}. The upper-left panel shows the mass distribution of stars in the \afe{}-\feh{} plane. Particle selections corresponding to "high-\afe{}", "low-\afe{}" and "high-\feh{}" are denoted by the overlaid red, blue and green boxes, respectively. Overlaid coloured tracks denote the evolution of the median abundances of the natal gas of these populations, with circle, square and triangle symbols corresponding to the epochs at which 25, 50 and 75 percent of the gas has been consumed, respectively. The evolution of the median \afe{} and \feh{} is plotted as a function of cosmic time in the centre-left and lower-left panels, respectively. Shaded regions on these panels denote the interquartile range. Dashed black tracks on the panels of the left-hand column denotes the SFR-weighted median gas-phase abundances. The upper-right panel shows the evolving median and interquartile range of the galactocentric radii (in proper coordinates) of the natal gas of each population. The centre-right panel shows the SFR-weighted mean consumption time of the natal gas, and the bottom-right panel shows the SFR-weighted mean of the natal gas mass fraction of Fe that was synthesised by Type Ia SNe. For these two panels, the shaded regions denote the $1\sigma$ scatter about the mean.}
\end{figure*}

The Lagrangian nature of the EAGLE simulations enables us to reconstruct the full enrichment history of the gas from which disc stars formed. It is therefore instructive to examine the evolution of the elemental abundances of the `natal' gas of the stars occupying key positions in the \afe{}-\feh{} plane at $z=0$. Although, as discussed in Section \ref{sec:subgrid_physics}, the mass and metals donated by stellar particles to SPH particles are `fixed' to particles and do not diffuse between them, any dilution of elemental abundances resulting from the inflow of low-metallicity gas can nonetheless since we examine kernel-smoothed abundances. In order to examine enrichment histories with superior temporal resolution to that afforded by the standard set of EAGLE snapshots, we focus here on the Ref-L025N0376 simulation, for which 1000 snapshots were recorded. One of the galaxies that forms in this simulation is, at $z=0$, a disc-dominated ($ \kappa_{\mathrm{co}} = 0.58$, $r_{1/2} = 8.8$ kpc) central galaxy that exhibits \afe{} bimodality at fixed \feh{}. Its stellar mass is $M_{*} = 4.58\times 10^{10}\mathrm{M_{\odot}}$, slightly below the lower bound of the mass interval used in Section \ref{sec:disc_stellar_pops} to define the Ref-L100N1504 sample, but this galaxy is a useful example owing to the similarity of its star formation history and elemental abundances with those of galaxy B. Its present-day halo mass is $M_{200} = 2.54\times 10^{12}\mathrm{M_{\odot}}$.

Fig. \ref{fig:daddyplot} summarises the enrichment history of this galaxy. The upper-left panel shows the \afe{}-\feh{} distribution of the galaxy's present-day disc stars as a 2-dimensional histogram. The overlaid boxes define three populations of stellar particles, for which we reconstruct the enrichment history of their natal gas particles. The boxes span $\Delta$\feh{}\,$=0.6$ and $\Delta$\afe{}\,$=0.1$, and the red and blue cases, respectively, correspond to high-\afe{} and low-\afe{} at the intermediate values of \feh{} for which the bimodality is most pronounced. The green case corresponds to the greatest values of \feh{}. The three samples comprise between approximately 400 and 900 stellar particles. The overlaid tracks of the same colour show the evolution, for each population, of the median \afe{} and \feh{} abundances of the gas particles that remain unconsumed by star formation. To avoid poor sampling of the measurement, the tracks truncate when only 30 gas particles remain unconsumed, and the evolution of the consumed fraction is shown via the symbols overlaid on each track, which denote the epochs at which 25 percent (circle), 50 percent (square) and 75 percent (triangle) of the gas particles have been consumed (these conditions apply to all tracks presented in this figure). Star formation does not uniformly sample the natal gas, so the elemental abundances of stars formed at a given epoch are not represented by the corresponding position on the median tracks; the dashed black track therefore shows the evolution of the SFR-weighted mean coordinate in \afe{}-\feh{} space. The tracks begin when at least 30 particles have a non-zero SFR. We stress that this track does not represent the enrichment history of any particular gas population, but rather the characteristic elemental abundances with which stars are being formed at the corresponding epoch.

It is immediately apparent that the enrichment histories of the natal gas of the high- and low-\afe{} populations are markedly different. They differ in \afe{} already at very low \feh{}, and the difference grows as the populations become more metal rich. The high-\feh{} population exhibits a similar enrichment history to the low-\afe{} population, but is offset to lower \afe{}; it is accreted at late times (as we shall discuss shortly, much of it is delivered by a gas-rich satellite) and enriches to high-\feh{} as it mixes with the interstellar gas of the evolved galaxy. The dashed black track shows that at early times, the $z=0$ disc stars of this galaxy initially form with elevated $\alpha$-elemental abundances and an increasing \feh{}, resulting in the formation of the high-\afe{} sequence. Once a little over 25 percent of all present day disc stars have formed, the formation of high-\afe{} stars begins to subside and the low-\afe{} sequence also begins to emerge. At this epoch the SFR-weighted mean track necessarily declines (and does so at roughly fixed \feh{}), and subsequently converges with the low-\afe{} sequence.

The clear and persistent separation of the median abundances of the natal gas of the high- and low-\afe{} populations signals that they never mixed. In contrast to a central assumption of the two-infall model, at no stage of its evolution does the natal gas of the low-\afe{} sequence reach values of \afe{} that are characteristic of the high-\afe{} population. Hence, rather than subsequently declining to roughly solar \afe{} in response to enrichment by Type Ia SNe ejecta, it simply never reaches values of \afe{} as elevated as those realised by the natal gas of the high-\afe{} population. 

The contrasting enrichment histories of the two gas populations are made more apparent by inspection of the centre-left and bottom-left panels of Fig. \ref{fig:daddyplot}. Respectively, these show the temporal evolution of \afe{} and \feh{}, with the thick solid lines denoting median values and the shaded regions the interquartile range. As with the upper-left panel, since star formation samples a very unrepresentative subset of the gas comprising each population, we stress that the coloured tracks do not denote the typical abundances with which the stars of each population form at the corresponding epoch. To give a sense of how the median abundances of each population differ from the latter, we again show the galaxy's SFR-weighted mean abundance as a dashed black track. All three gas populations do initially become $\alpha$-enriched in response to early star formation; the natal gas of the high-\afe{} stars is rapidly enriched to \afe{}$\,\simeq 0.5$ and \feh{}$\,\simeq -0.5$, and is mostly consumed by $t\simeq 4\,{\rm Gyr}$. After its initial enrichment with $\alpha$-elements, the low-\afe{} population's natal gas settles to a value of \afe{}$\,\simeq 0.35$ at $t\simeq 5\,{\rm Gyr}$ ($z \simeq 1.5$), after which it is steadily enriched by Type Ia SNe as it is consumed. This gradually reduces its \afe{}, and increases its \feh{}, in a broadly monotonic fashion. The evolution of \feh{} for this population is much more gradual than is the case for the natal gas of the high-\afe{} stars, the latter reaching \feh{}$\,\gtrsim -0.5$ more than $3\,{\rm Gyr}$ sooner than the former. The correspondence of the high-\feh{} population with the low-\afe{} population is also apparent from inspection of the time evolution of \afe{} and \feh{}. The high-\feh{} population represents the late-infalling subset of the overall low-\afe{} sequence, resulting in a lower abundance of $\alpha$ elements and a higher abundance of Fe. 

We can conclude from analysis of this galaxy that the elemental abundances of the natal gas of its high- and low-\afe{} populations evolved along distinct paths. Although the latter briefly exhibits elevated \afe{} at early times, which subsequently declines in response to enrichment by Type Ia SNe, this gas never reaches \afe{} comparable to that of the stars comprising the high-\afe{} sequence. The formation of the high-\afe{} sequence is therefore not imprinted in the abundances of the gas from which the low-\afe{} sequence subsequently forms, as one might expect if, as postulated by the two-infall model, the low-\afe{} sequence forms from interstellar gas remaining after an initial episode of star formation. This notwithstanding, the characteristic $\alpha$-element and Fe abundances with which stars form, quantified by the mean SFR-weighted coordinate in \afe{}-\feh{} space, does evolve in a fashion that is qualitatively similar to the expectations of the two-infall model: the typical \afe{} declines rapidly as the formation of the high-\afe{} sequence draws to a close and the low-\afe{} sequence begins to dominate. However, as was also noted recently by \citet{2017arXiv170807834G} following analysis of the Auriga simulations, the continuous increase of \feh{} of the natal gas of all populations is incompatible with a key expectation of the two-infall model, namely that the metallicity of the gas from which the Milky Way's disc stars are formed converges towards an equilibrium value.

The dissimilar evolutionary histories of the natal gas of the stars comprising high- and low-\afe{} sequences implies that these gas populations remained physically separated, and hence chemically independent, prior to their consumption by star formation. This suggests that they accreted onto the galaxy at different times. To explore this possibility, we crudely reconstruct the collapse history of the natal gas of the three populations defined in the upper-left panel of Fig. \ref{fig:daddyplot}. This is achieved by computing the spherical galactocentric radii of the unconsumed gas particles as a function of cosmic time, relative to the coordinate of the most-bound particle of the galaxy's main progenitor subhalo. The resulting trajectories are shown in the upper-right panel, with the tracks denoting the median radius and the shading denoting the interquartile range. The plot adopts proper (rather than comoving) coordinates, to highlight the expansion of gas with the Hubble flow at epochs prior to turnaround, and the horizontal dotted line at $30\,{\rm pkpc}$ provides a threshold for considering the gas to have accreted onto the galaxy; this definition is consistent with that used in Section \ref{sec:finding_galaxies}. We again stress that, since star formation does not uniformly sample the gas comprising each population, the median tracks do not denote the typical radius at which stars form at each epoch; we do not include a SFR-weighted track here, as the value is always $\ll 30\,{\rm pkpc}$. 

The high-\afe{} population reaches its radius of maximum expansion (the "turnaround radius") earlier ($t\simeq 1\,{\rm Gyr}$) than the low-\afe{} population ($t\simeq 2\,{\rm Gyr}$), and does so at a median galactocentric radius that is more than a factor of two smaller. A significant fraction of natal gas of the low-\afe{} population is delivered by a gas-rich satellite at $t\simeq 10\,{\rm Gyr}$, which induces the oscillatory structure in its median radius. The natal gas of the high- and low-\afe{} populations is in general not co-spatial, precluding significant mixing of their kernel-smoothed element abundances: the median galactocentric radius of the high-\afe{} population drops below $30\,{\rm pkpc}$ at $t=3.4\,{\rm Gyr}$, at which time only 3 percent of the particles comprising the low-\afe{} population are located within $30\,{\rm pkpc}$. The median galactocentric radius of the latter falls below $30\,{\rm pkpc}$ much later, at $t\simeq 9\,{\rm Gyr}$, by which time all of the high-\afe{} population's stars have already formed. 

The influence of the accretion history on the enrichment of these populations is made clear by the centre-right and bottom-right panels of Fig. \ref{fig:daddyplot}. The former shows the evolution of the consumption timescale, $t_{\rm g}$, of the gas, the latter shows the evolution of the gas-phase mass fraction of Fe synthesised by Type Ia SNe $f_{\rm Fe,SNIa}$. Here, $t_{\rm g}$ is computed as per Equation \ref{eq:t_g}, replacing $P_\star$ by the gas particle's pressure. For consistency with Fig. \ref{fig:afetcon}, we assume this to be the equation of state pressure corresponding to the density of the gas, $P_{\rm eos}(\rho)$. Since we are specifically concerned in these two panels with the subset of the gas sampled by star formation, the coloured tracks here show the SFR-weighted mean of the quantity in question, and they begin when at least 30 particles have a non-zero SFR. The shaded regions show the $16^{\rm th}-84^{\rm th}$ percentile scatter, as an estimate of the $1\sigma$ scatter about the mean. 

The early collapse of the natal gas of the high-\afe{} population drives it to high densities and pressures, fostering its rapid conversion to stellar particles with a characteristic consumption timescale that is typically $t_{\rm g} \lesssim 500\,{\rm Myr}$, and hence much shorter than the e-folding ($\tau=2\,{\rm Gyr}$, upper dashed line) and half-life ($t_{1/2}=\tau\log(2)=1.39\,{\rm Gyr}$, lower dotted line) timescales of the Type Ia SNe delay time function. In contrast, over 75 percent of the stars comprising the low-\afe{} sequence (and nearly all of the high-\feh{} stars) form from gas with a consumption timescale of $t_{\rm g} \gtrsim 1.0\,{\rm Gyr}$. The influence of this dichotomy on the enrichment of star-forming gas by Type Ia SNe is then clear from inspection of the bottom-right panel of Fig \ref{fig:daddyplot}. At early epochs, when the high-\afe{} stars form, the typical fraction is $f_{\rm Fe,SNIa} \simeq 0.35$. At $t \gtrsim 4\,{\rm Gyr}$ there is a clear jump in the typical Fe mass fraction from Type Ia SNe, with the low-\afe{} and high-\feh{} populations, which largely form after this transition, typically exhibiting $f_{\rm Fe,SNIa} \simeq 0.55$.

\begin{figure}
\includegraphics[width=\columnwidth]{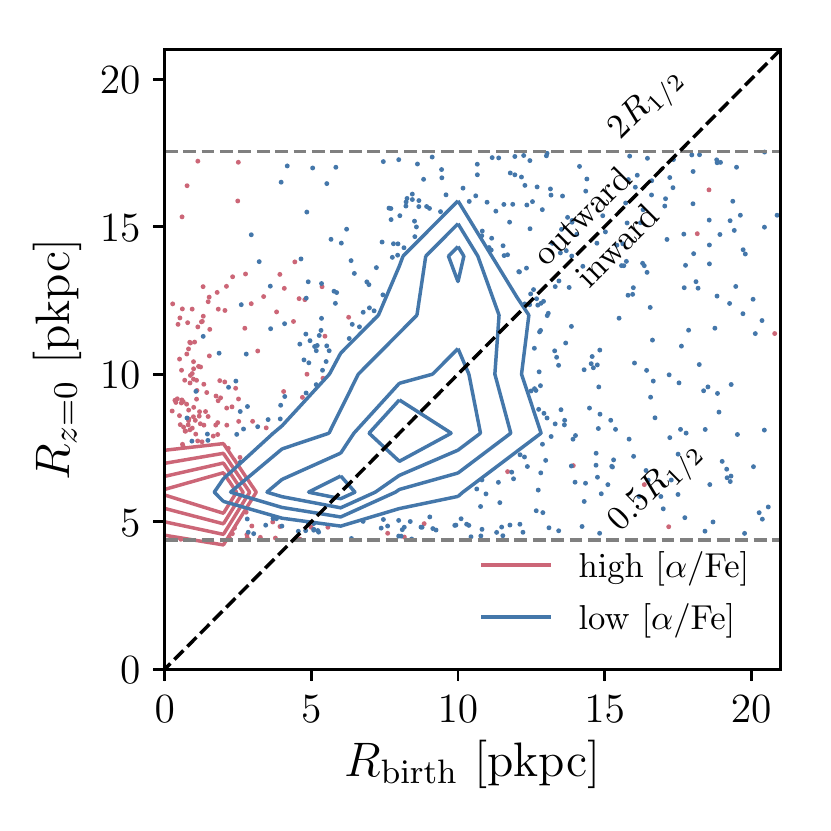}
\caption{\label{fig:birth_vs_z0_radii} The cylindrical radii at the present day, $R_{z=0}$, of the stellar particles comprising the high- (red) and low-\afe{} (blue) populations of the example galaxy from Ref-L025N0376, as a function of their cylindrical radii at birth, $R_{\rm birth}$. The distribution is shown with contours denoting the levels containing 50, 60, 70 and 80\% of the points, with individual particles drawn beyond the outer (80\%) contour. Grey dashed lines at $R_{z=0} = 0.5R_{1/2}$ and $R_{z=0} = 2R_{1/2}$ show the cylindrical radial boundaries used to define the disc stars, and the black dashed line denotes the locus $R_{z=0} = R_{\rm birth}$. The low-\afe{} population has experienced mild outward radial migration, albeit with large scatter. The high-\afe{} population has necessarily experienced significant outward radial migration in order to be identified as part of the disc, since these stars formed almost exclusively within $5\,{\rm pkpc}$ of the galactic centre.}
\end{figure}

Returning briefly to the upper-right panel of Fig. \ref{fig:daddyplot}, it is striking that the gas from which the stars comprising the high-\afe{} sequence formed was significantly more compact prior to its consumption than was the gas that fuelled the formation of the low-\afe{} stars. This is perhaps to be expected; linear tidal torque theory \citep{1984ApJ...286...38W,1996MNRAS.282..455C} posits that the angular momentum of matter grows whilst it expands with the Hubble flow ($L \propto a^{3/2}$), and remains constant after turnaround. Gas that reaches its radius of maximum expansion later is therefore expected to build more angular momentum, and settle farther out in the galaxy disc. This raises the question of how the stars of both the high- and low-\afe{} populations came to be broadly co-spatial at the present day. 

Fig. \ref{fig:birth_vs_z0_radii} plots the distribution of the present-day cylindrical galactocentric radius of the high- and low-\afe{} stars, $R_{\rm z=0}$, of the example galaxy from Ref-L025N0376, as a function of their cylindrical galactocentric radius at birth, $R_{\rm birth}$. Horizontal dashed lines denote $0.5R_{1/2}$ and $2R_{1/2}$, the cylindrical radial boundaries we impose to select disc stars at $z=0$, whilst the black dashed line denotes the locus $R_{z=0} = R_{\rm birth}$. The low-\afe{} population generally tracks this locus, albeit with a large scatter such that significant positive and negative migration is common. Overall there is a mild preference for net positive migration. In contrast, the high-\afe{} population of this galaxy forms almost entirely at $R_{\rm birth} < 3\,{\rm pkpc}$, which is likely a necessary condition in order to realise the short consumption timescales that preclude the enrichment of their natal gas with Type Ia SNe ejecta. Positive outward migration of these stellar particles is therefore necessary in order for them to be categorised as part of the (geometric) disc. 

Isolating the physical cause of this migration is beyond the scope of this study. We note that the limited resolution of our simulations, and their implicit treatment of the cold interstellar gas phase, preclude an examination of whether such migration might be dominated by the `blurring' or `churning' processes discussed by \citet{2009MNRAS.396..203S}. That notwithstanding, as recently argued by \citet{2017arXiv170901040N} following analysis of the high-resolution APOSTLE simulations \citep{2016MNRAS.457.1931S} that also use the EAGLE galaxy formation model, dynamical effects similar to those expected to follow from detailed internal processes can also arise in response to the evolution of the ISM from a `thick' state at early times to a more settled state at low redshift, as the accretion of gas onto (and ejection of gas from) galaxy discs declines \citep[see also][]{2004ApJ...612..894B,2012MNRAS.426..690B,2013ApJ...773...43B,2016arXiv160804133M}.

\begin{figure*}
\includegraphics[width=1.\textwidth]{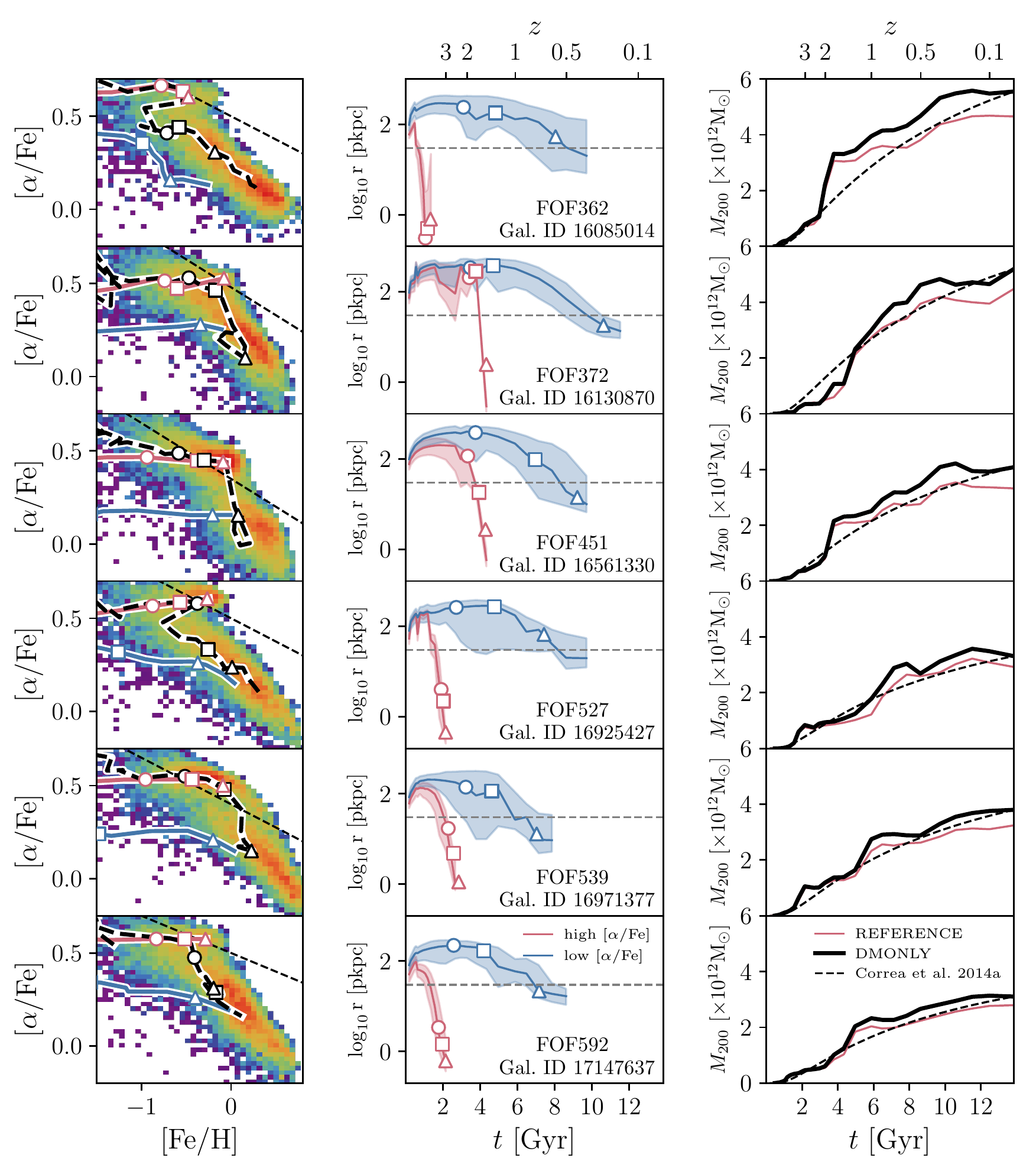}
\caption{\label{fig:bimodalexamples} The evolution of the six galaxies from Ref-L100N1504 that exhibit bimodality in \afe{} at fixed \feh{}. Similarly to Fig. \ref{fig:daddyplot}, the left-hand column shows the \afe{}-\feh{} distribution of the disc stars of these galaxies as a 2-dimensional histogram. Here the disc stars are split into high- and low-\afe{} populations with a simple visually-defined cut, shown by the thin dashed line. The centre column shows the expansion, collapse and accretion onto the galaxy of the natal gas of these two populations. The right-hand column, discussed in Section \ref{sec:halo_accretion}, shows the total mass accretion history of the galaxies, i.e. $M_{200}(t)$. The red curve shows this quantity within the Ref-L100N1504 simulation, whilst the black curve denote the evolution of the same halo identified in its dark matter-only counterpart, DMONLY-L100N1504. Here, the dashed black curve denotes the typical accretion history of haloes with the same present-day mass as the DMONLY realisation of the halo, as parametrised by \citet{2015MNRAS.450.1514C}.} 
\end{figure*}

\subsection{Applicability to galaxies in the Ref-L100N1504 simulation}
\label{sec:broader_applicability}

Visual inspection of the \afe{}-\feh{} planes of the 133 EAGLE galaxies in our sample reveals six that exhibit two clearly-separated sequences in \afe{}-\feh{} space. The \afe{}-\feh{} distributions of these galaxies are shown in the left-hand column of Fig. \ref{fig:bimodalexamples}. In a similar fashion to the case discussed in Section~\ref{sec:afeorigin}, we split the population of disc stars of these galaxies into high- and low-\afe{} sequences, to examine their evolution separately. These examples exhibit a diversity of \afe{}-\feh{} morphologies, so we adopt simple, visually-defined cuts that assign all disc stars to one of the two sequences, as indicated by the dashed lines on each panel. The cuts are defined such that they best separate the sequences where they are well separated. The overlaid tracks and symbols are defined as per the top-left panel of Fig. \ref{fig:daddyplot}. Similarly, the centre column shows the collapse histories of the gas from which the two populations formed, as per the upper-right panel of Fig. \ref{fig:daddyplot}. We discuss the right-hand column later in Section \ref{sec:halo_accretion}.

These panels reveal illuminating features, similarities and differences with respect to the example in Fig. \ref{fig:daddyplot}. As there, the natal gas populations of the stars comprising the high- and low-\afe{} sequences exhibit distinct enrichment histories, implying a general lack of mixing of their element abundances prior to their consumption by star formation. In all cases, the natal gas of the high-\afe{} sequence reaches a relatively short radius of maximum expansion, "turning around" and accreting onto the galaxy significantly earlier than that of the low-\afe{} sequence. This early collapse enables rapid formation of the high-\afe{} sequence, with the short consumption timescales necessary to inhibit enrichment by the ejecta of Type Ia SNe. As might be expected, the most $\alpha$-rich sequences are formed by the galaxies for which the natal gas collapses at the earliest epochs (FOF362, FOF527 and FOF592), whilst the example for which the natal gas of the two sequences is the most cospatial prior to its accretion onto the galaxy (FOF372) exhibits the least well-separated sequences in \afe{}-\feh{} space. The galaxies also exhibit diversity in the maximum \feh{} abundance to which the high-\afe{} sequence extends; as expected, we find this is correlated to the mass of stars formed from the gas delivered in the initial episode.

The track of the mean SFR-weighted element abundances exhibits significant diversity, reflecting the non-trivial enrichment evolution of the \emph{star-forming} ISM. The tracks for galaxies FOF 362 and FOF527 illustrate that, as the formation of the high-\afe{} sequence concludes and the formation of low-\afe{} stars begins to dominate, the decline of the characteristic \afe{} is accompanied by a temporary decrease in the characteristic Fe abundance with which new stars are formed (by $\Delta\mathrm{[Fe/H]}\lesssim -0.5$). This indicates that, as posited by the two-infall model, in these two cases the accretion of unenriched gas temporarily lowers the metallicity of the star-forming ISM. Conversely, in the other four bimodal galaxies, the decrease in \afe{} associated with the transition from the high- to low-\afe{} sequence is accompanied by mild increase in \feh{}. 

\section{Connecting disc star element abundances to halo mass accretion histories}
\label{sec:halo_accretion}

The analyses presented in Section \ref{sec:afeorigin} show that distinct sequences in \afe{}-\feh{} space arise in galaxies where disc stars are formed from gas that was accreted onto the galaxy in somewhat separated episodes. Since the accretion of gas is primarily driven by the gravitational evolution of a galaxy's dark matter halo \citep{2008MNRAS.388.1792N,2010MNRAS.406.2267F,2015MNRAS.450.1521C,2015MNRAS.450.1514C,2015MNRAS.452.1217C}, this finding highlights that the distribution of disc stars in the \afe{}-\feh{} plane is influenced not only by astrophysical processes, but perhaps more fundamentally by the hierarchical formation and assembly of the galaxy. We therefore briefly examine in this section the connection between bimodality in \afe{} at fixed \feh{}, and the accretion history of a galaxy's dark matter halo.

The right-hand column of Fig. \ref{fig:bimodalexamples} shows the total mass accretion history of the six bimodal galaxies discussed in Section \ref{sec:broader_applicability}, characterised by the evolution of the halo virial mass, $M_{200}(t)$. In each case, the red curve shows growth of the virial mass of the halo within the Ref-L100N1504 simulation. The thick solid curve shows the evolution of $M_{200}^{\rm DMO}(t)$, the virial mass of the same halo, identified using the techniques described by \citet{2015MNRAS.453L..58S}, in a simulation of the L100N1504 volume considering only collisionless gravitational dynamics (DMONLY-L100N1504). This track is instructive as it eliminates the influence of astrophysical processes associated with baryons, such as adiabatic contraction, which acts to increase the central density of the halo, and the reduction of the halo's mass and accretion rate due to the ejection of baryons by feedback processes. The dashed curve shows the typical accretion history of a halo with the same present-day mass as the halo in the DMONLY-L100N1504 simulation, as derived by \citet{2015MNRAS.450.1514C} through application of the extended Press-Schechter (EPS) formalism to the growth rate of initial density perturbations.

The importance of accounting for the influence of baryon physics, in particular ejective feedback, is made clear by comparison of the Reference and DMONLY curves. In each case, these curves diverge significantly at intermediate redshifts ($1 \lesssim z \lesssim 3$). As shown by \citet[][their Fig. 1]{2015MNRAS.453L..58S}, the median of the ratio $M_{200}/M_{200}^{\rm DMO}$ at $z=0$ is $\simeq 0.85$ for haloes of $M_{200}^{\rm DMO} \sim 10^{12}-10^{13}\,{\rm M}_\odot$  Comparison of the DMONLY curves with the \citet{2015MNRAS.450.1514C} fitting function reveals that a common feature of the mass accretion histories of the 6 bimodal galaxies is rapid growth at early times ($4 \lesssim t \lesssim 6\,{\rm Gyr}$ or $1 \lesssim z \lesssim 2$), such that early in their assembly, they grow faster than is typical for haloes of the same present-day mass. Their accretion rate subsequently declines relative to the average rate for haloes of the same z=0 mass, such that at late times $M_{200}^{\rm DMO}(t)$ grows at a significantly \textit{lower} rate than is typical. Na\"ively, an elevated mass accretion rate at early cosmic epochs is qualitatively consistent with the early collapse of the gas that fuels the formation of the high-\afe{} sequence. This also motivates the question of whether the haloes that host disc-dominated galaxies with distinct sequences in \afe{}-\feh{} space differ significantly and systematically from those hosting disc-dominated galaxies with unimodal \afe{}-\feh{} distributions. 

\begin{figure}
\includegraphics[width=\columnwidth]{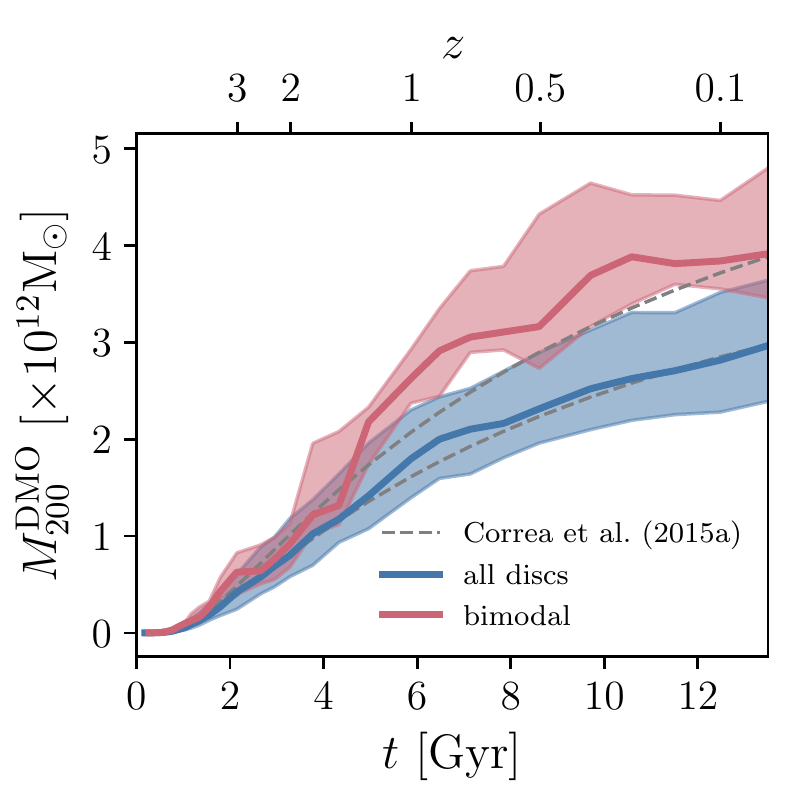}
\caption{\label{fig:comp_uni_bi_acc_hist} The mass accretion history, $M_{200}(t)$, of haloes identified in the DMONLY-L100N1504 simulation. The thick curves corresponds to the median, whilst the shaded regions denote the interquartile range. The blue curve corresponds to the partner haloes of all 133 galaxies in our Milky Way-like sample, whilst the red curve corresponds to the partner haloes of the 6 galaxies exhibiting bimodality in \afe{} at fixed \feh{}. Dashed curves meeting the two median curves at $z=0$ show the typical accretion history, as parametrised by \citep{2015MNRAS.450.1514C}, of haloes with these $z=0$ masses. Haloes which host galaxies exhibiting \afe{} bimodality have systematically different dark matter accretion histories}
\end{figure}

We therefore identify the DMONLY counterpart haloes of all 133 Milky Way-like galaxies in Ref-L100N1504. We plot in Fig. \ref{fig:comp_uni_bi_acc_hist} the median and interquartile range of $M_{200}^{\rm DMO}(t)$ derived from these samples (blue curves denote the entire sample, red curves denote the six bimodal galaxies). As per the right-hand column of Fig. \ref{fig:bimodalexamples}, the overlaid dashed curves that intersect the median curves at $z=0$ denote the typical accretion histories, as parametrised by Correa et al. (2017a), of haloes with the same present-day mass. Comparison of the two median tracks indicates that the phase of rapid growth exhibited by bimodal galaxies between $t \simeq 4$ and $t \simeq 8\,{\rm Gyr}$ (corresponding to $z \simeq 1.6$ and $z \simeq 0.5$) is not common amongst the broader population of Milky Way-like galaxies. In general, the latter exhibit accretion histories that are much more representative of the entire population of haloes with similar present-day mass. We therefore conclude that the accretion histories of bimodal galaxies are indeed atypical. We note that the present day halo masses of the bimodal galaxies are not representative of the overall distribution; the median $M_{200}^{\rm DMO}(z=0)$ of the bimodal galaxies is equal to the $80^{\rm th}$ percentile of that of the overall sample of Milky Way-like galaxies. However, we have verified that median accretion history of the latter remains largely unchanged if one examines instead a sub-sample of these haloes whose present day masses span the same range as those of the bimodal galaxies.

An early and rapid phase of mass accretion onto a galaxy's halo therefore appears to be a necessary (but perhaps not sufficient) condition for the emergence of distinct sequences in the \afe{}-\feh{} distribution of disc stars. Further, more detailed examination of the connection between halo accretion histories and the elemental abundances of disc stars is beyond the scope of this study, however we can already note two implications of these findings. Firstly, since halo accretion histories are purely a consequence of the initial phase-space configuration of the matter destined to comprise a galaxy's dark matter halo, whether or not a galaxy will develop bimodal sequences in the \afe{}-\feh{} plane is effectively determined at early cosmic epochs. Secondly, the key role played by a galaxy's accretion history implies that predictive modelling of the emergence of the \afe{}-\feh{} distribution of disc stars requires that the formation and assembly of the galaxy is considered in its cosmological context. Specifically, this entails accounting for growth, merging history and chemical evolution of a galaxy's progenitors. These processes are incorporated self-consistently in cosmological hydrodynamical simulations, but are challenging to incorporate realistically into analytic models. 

\section{Summary and discussion}
\label{sec:summary_and_discussion}

We have examined the enrichment history of disc stars in present-day galaxies in the EAGLE simulations of galaxy formation. In particular, we have focussed on the formation and assembly of galaxies whose disc stars exhibit distinct sequences in the \afe{}-\feh{} plane. Our findings can be summarised as follows:

\begin{itemize}

\item The distribution in \afe{}-\feh{} space of the disc stars of simulated Milky Way-like galaxies
is characterised by a roughly constant $\alpha$-element abundance of \afe{}$\simeq 0.5$ for \feh{}$< -0.5$, whilst it declines roughly linearly at higher Fe abundance, to \afe{}$\simeq -0.1$ at \feh{}$=1.0$ (Fig. \ref{fig:afestack}).

\item At any fixed \feh{}, \afe{} anti-correlates linearly with $f_{\rm Fe,SNIa}$, the mass fraction of a star's Fe that was synthesised by Type Ia SNe. This demonstrates that the key driver of elevated $\alpha$-element abundances is, unsurprisingly, the smaller fraction of Fe synthesised by this channel. There is also a weaker trend such that at fixed \afe{}, $f_{\rm Fe,SNIa}$ declines with increasing \feh{}, since the main pathway for increasing \feh{} without reducing \afe{} is to source a greater fraction of Fe from the same source as the $\alpha$ elements, i.e. Type II SNe (Fig. \ref{fig:afesniafrac}).

\item Broadly, the oldest stars exhibit low \feh{} and high \afe{}, whilst the youngest stars exhibit the opposite. But the correlation of age with both abundance diagnostics saturates such that large areas of the \afe{}-\feh{} plane exhibit similar characteristic ages, precluding the use of either as an accurate chronometer and demonstrating that age is not the sole driver of element abundances (Fig. \ref{fig:afeages}).

\item At fixed \feh{}, the most $\alpha$-rich stars formed from gas with the shortest consumption timescales, highlighting that the key to yielding high values of \afe{} is the consumption of star-forming gas before it can be substantially enriched with Fe from Type Ia SNe (Fig. \ref{fig:afetcon}).

\item The distributions of the disc stars of rotationally-supported galaxies with similar mass to the Milky Way in the \afe{}-\feh{} plane are diverse. The majority are unlike that of the Milky Way, insofar as they do not exhibit two distinct sequences. Only $\simeq 5$ percent of Milky Way-like galaxies in EAGLE exhibit bimodality in \afe{} at fixed \feh{}. A few galaxies exhibit only the high-\afe{} sequence. The distribution is closely connected to the star formation history of the disc stars, with the high-\afe{} sequence resulting from intense star formation at early times ($z \gtrsim 2$), and the low-\afe{} sequence from extended star formation  at later times (Fig. \ref{fig:afeexamples}).

\item In galaxies exhibiting distinct high- and low-\afe{} sequences, the gas from which the stars comprising each sequence formed is accreted onto the galaxy in distinct episodes. This temporal separation inhibits mixing of the two gas populations, enabling divergent evolution of their element abundances from an early epoch. The low-\afe{} sequence does not form (primarily) from interstellar gas left unconsumed by the formation of the high-\afe{} sequence, therefore the oldest stars of the former can exhibit lower \feh{} than the youngest stars of the latter. The median Fe abundance of both sequences increases monotonically and continuously and, in contrast to the common assumption of analytic models, does not reach a constant equilibrium value (Fig. \ref{fig:daddyplot}).

\item The early collapse of the natal gas of the high-\afe{} stars fosters star formation with short consumption times, precluding strong enrichment by Type Ia SNe (Fig. \ref{fig:daddyplot}).

\item The formation of high-\afe{} stars from gas with short consumption timescales requires that they are born in a compact configuration; they are found in the disc at the present day having experienced a net outward radial migration over $\simeq 8-10\,{\rm Gyr}$ (Fig. \ref{fig:birth_vs_z0_radii}).

\item We identify six galaxies from our sample of 133 ($\simeq 5$ percent), whose disc stars exhibit distinct sequences in \afe{}-\feh{} space. In each case, the formation of the high-\afe{} sequence is associated with the early infall of gas onto the galaxy, which is rapidly consumed by star formation. In galaxies for which the two accretion episodes are more clearly separated, the two sequences are also more distinct (Fig. \ref{fig:bimodalexamples}).

\item The dark matter haloes that host the six bimodal galaxies exhibit (total) mass accretion histories characterised by a rapid phase of growth at intermediate epochs ($1 \lesssim z \lesssim 3$), followed by a tailing off to a significantly lower rate of growth. Such accretion histories are atypical, as highlighted by comparison with the average accretion history of dark matter haloes with the same present-day virial mass (Fig. \ref{fig:bimodalexamples}). Milky Way-like galaxies that exhibit only the low-\afe{} sequence have dark matter halo accretion histories that are much more typical (Fig. \ref{fig:comp_uni_bi_acc_hist}). 

\end{itemize}

The results presented here demonstrate that realistic cosmological hydrodynamical simulations do form galaxies whose present-day disc stars exhibit two distinct sequences in the \afe{}-\feh{} plane, as revealed by spectroscopic surveys of the Galaxy. 
As also shown by \citet{2017arXiv170807834G}, such sequences form in response to distinct episodes of gas accretion onto galaxies. We have shown that distinct accretion episodes lead to the formation of stars with differing characteristic gas consumption timescales: whilst the low-\afe{} sequence forms from gas whose consumption timescale is similar to the e-folding timescale of the Type Ia SNe delay time distribution, the natal gas of the high-\afe{} sequence is consumed on a much shorter timescale, suppressing enrichment by Type Ia SNe.

Our results corroborate the conclusion of \citet{2017arXiv170807834G} that such a dichotomy is not ubiquitous\footnote{The initial conditions of the Auriga simulations are drawn from the EAGLE Ref-L100N1504 volume; The EAGLE galaxies that form in the same haloes resimulated for the Auriga Project do not satisfy our selection criteria (Section \ref{sec:finding_galaxies}), but we have examined these galaxies and found their \afe{}-\feh{} planes to be qualitatively similar to those presented by \citet{2017arXiv170807834G}.} \citep[c.f also][]{2012MNRAS.426..690B,2014A&A...572A..92M}. The large volume of the EAGLE Ref-L100N1504 simulation yields a sample of 133 `Milky Way-like' galaxies (defined on the basis of their stellar mass and kinematics), enabling us to place, for the first time, galaxies with distinct \afe{}-\feh{} sequences into the broader context of the galaxy population. The relative scarcity within EAGLE of galaxies with \afe{}-\feh{} distributions similar to that observed in the Galaxy indicates that in this respect the Milky Way is likely unrepresentative of the broader population of $\sim L^\star$ late-type galaxies. We have demonstrated that such abundance patterns are likely to prove uncommon because distinct gas accretion episodes require an atypical mass accretion history, characterised by a phase of rapid growth at relatively early epochs. That the Galaxy's element abundances and dark matter halo accretion history may be unrepresentative of the broader population of similarly massive disc galaxies suggests that caution should be exercised when generalising the findings of Galactic surveys, particularly for `near-field cosmology' applications.

The conclusion of \citet{2017arXiv170807834G}, that the enrichment history of galaxies in cosmological simulations contrasts with the expectations of leading analytic Galactic chemical evolution models, is also corroborated by our findings. The EAGLE simulations indicate that the metallicity of the natal gas of disc stars tends to increase continuously over time, and does not tend to an equilibrium value established by the balance of enrichment and gas infall. Our conclusion that two distinct accretion episodes are necessary to foster a bimodality in \afe{} is an aspect in common with the \citet{1997ApJ...477..765C,2001ApJ...554.1044C} two-infall model, but in other respects the simulations differ markedly from that model. The major difference is that the latter's assumption of instantaneous and complete mixing of star-forming gas precludes the contemporaneous formation of stars with high- and low-\afe{}, and hence unavoidably imprints the natal gas of the low-\afe{} sequence with a chemical record of the formation of the high-\afe{} sequence. By contrast, although our simulations indicate that high-\afe{} stars do in general form prior to their low-\afe{} counterparts, the age distributions of the two populations can and do overlap. This is possible because the populations of gas from which they form remain largely unmixed prior to their consumption, precluding significant mixing of their element abundances. 

A systematic uncertainty to which our simulations are subject, and whose influence we are unable to test directly, is the potential for greater mixing of heavy elements in real galaxies than is realised by the EAGLE simulations. As shown by \citet{2016arXiv160406803C}, relatively little interstellar gas ejected from EAGLE galaxies is later reincorporated into the ISM to form stars. Other galaxy formation simulations have reported more prevalent reincorporation of ejected gas \citep[see, e.g.][]{2017MNRAS.470.4698A}, and this process remains poorly-constrained by observations. Greater recycling of gas within the circumgalactic medium (CGM) can in principle promote the mixing of gas that fuels the formation of disc stars, inhibiting the formation of very distinct sequences in \afe{}-\feh{} space. However, since in the majority of cases examined here the natal gas of the high-\afe{} sequence is consumed prior to the gas of the low-\afe{} sequence reaching the turnaround radius, significant mixing is likely precluded, irrespective of the prevalence of halo recycling. Nonetheless, should mixing be found to be underestimated in the EAGLE simulations, the likely conclusion is that galaxies with element abundances similar to the Milky Way are even more rare than inferred here.

The recent advent of large cosmological simulations that reproduce a broad array of observed galaxy properties presents an exciting opportunity to advance studies of the Galaxy's chemical history using forward modelling, in contrast to the largely inverse methods that attempt to reverse engineer its observed elemental abundances. In seeking to further our understanding of the formation and assembly of the Milky Way, and galactic archaeology more generally, we expect on-going improvements to the detail and realism of cosmological simulations to prove an important ally to the development of next generation of spectroscopic surveys of the Galaxy.

\section*{Acknowledgements}
We thank Joel Pfeffer for assistance with the creation of galaxy merger trees and Marie Martig for helpful comments on an earlier version of the manuscript. JTM acknowledges an STFC doctoral studentship. RAC is a Royal Society University Research Fellow. Analyses and plots presented in this article used \texttt{iPython}, and packages in the \texttt{SciPy} ecosystem \citep{Jones:2001aa,4160265,4160251,5725236}. The study made use of high performance computing facilities at Liverpool John Moores University, partly funded by the Royal Society and LJMU's Faculty of Engineering and Technology, and the DiRAC Data Centric system at Durham University, 
operated by the Institute for Computational Cosmology on behalf of the 
STFC DiRAC HPC Facility (www.dirac.ac.uk). This equipment was funded by 
BIS National E-infrastructure capital grant ST/K00042X/1, STFC capital 
grants ST/H008519/1 and ST/K00087X/1, STFC DiRAC Operations grant 
ST/K003267/1 and Durham University. DiRAC is part of the National 
E-Infrastructure. This work was supported by the Netherlands Organisation for Scientific Research (NWO), through VICI grant 639.043.409. 



\bibliographystyle{mnras}
\bibliography{bib} 




 \begin{appendix}
 \section{Subgrid physics variations}
 \label{sec:subgrid}

\begin{figure*}
\includegraphics[width=1.\textwidth]{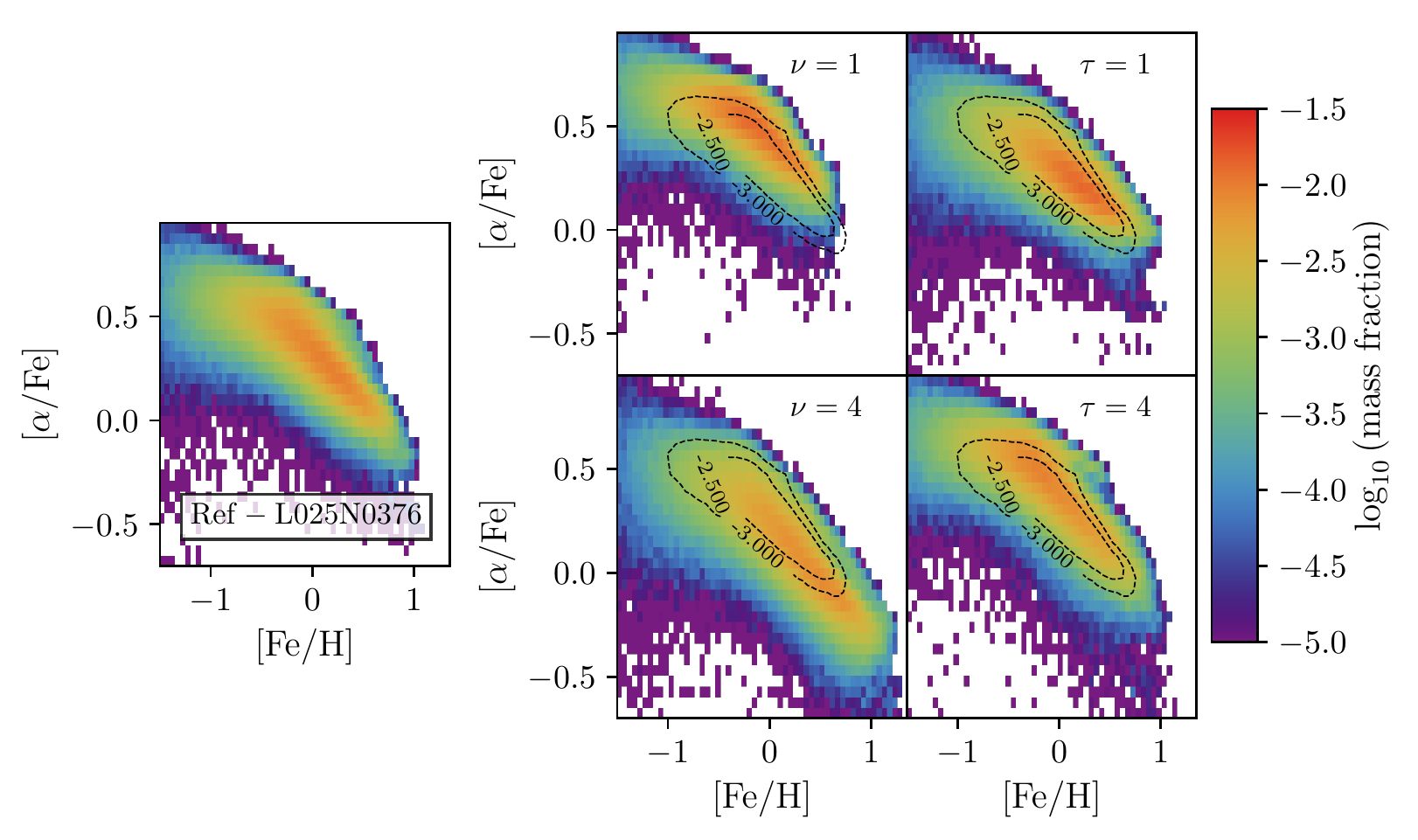}
\caption{\label{fig:subgrid} The \afe{}-\feh{} distribution of Milky Way like galaxies in simulations of the L025N0376 volume. The left panel shows the distribution realised by the Ref-L025N0376 simulation, whilst the four panels on the right show that from simulations in which a parameter governing the number of Type SNIa per unit stellar mass formed, $\nu$, or the characteristic e-folding timescale of the Type SNIa delay function, $\tau$, has been varied. On these panels, the overlaid black contours are from the Ref-L025N0376 distribution, highlighting the significant changes to the \afe{}-\feh{} distribution induced by these parameter changes.}
\end{figure*}

We briefly examine in this appendix the degree to which variation of the subgrid parameters governing the rate of Type Ia SNe influences the \afe{}-\feh{} distribution of disc stars of Milky Way-like galaxies. We analyse four simulations that adopt the same initial conditions as the Ref-L025N0376 simulation, two of which vary the total number of Type Ia SNe per unit of initial stellar mass formed, adopting $\nu = 1\times 10^{-3}\,{\rm M}_{\odot}^{-1}$ and $\nu = 4\times 10^{-3}\,{\rm M}_{\odot}^{-1}$ (relative to the Reference model, which adopts $\nu = 2\times 10^{-3}\,{\rm M}_{\odot}^{-1}$) , and two of which vary the characteristic e-folding timescale of the Type Ia SNe delay time distribution, adopting $\tau = 1\,{\rm Gyr}$ and $\tau = 4\,{\rm Gyr}$ (where the Reference model assumes $\tau = 2\,{\rm Gyr}$).

We examine the 56 galaxies in the Ref-L025N0376 simulation with $\kappa_\mathrm{co} > 0.4$ and stellar mass in the interval $4 < M_* < 8\times10^{10}\ \mathrm{M_{\odot}}$. We identify the same haloes in the 4 variation runs using the same particle matching technique \citep{2015MNRAS.453L..58S} used to pair haloes with their counterpart in the DMONLY simulation.

The right four panels of Fig. \ref{fig:subgrid} show the \afe{}-\feh{} distribution of these galaxies in the varied simulations as 2-dimensional histograms, with the overlaid contours showing the equivalent distribution from the Reference simulation, shown in its entirely in the left panel. Varying the number of Type Ia SNe per unit stellar mass formed has a significant impact on the distribution, as changing the number of Type Ia SNe also changes the total mass of Fe synthesised per unit stellar mass formed. Decreasing (increasing) the number of Type Ia SNe therefore shifts the distribution upward (downward) to a higher (lower) \afe{}, and truncates the distribution at a lower (higher) \feh{}. The characteristic delay timescale of Type Ia SNe governs the likelihood of gas becoming enriched with the Fe synthesised by Type Ia SNe whose progenitors formed recently. A shorter (longer) delay timescale results in a greater (lesser) fraction of stars forming from Fe-rich gas, inhibiting (aiding) the formation of a high-\afe{} sequence. 

These results indicate that the parameters governing the subgrid implementation of enrichment by Type Ia SNe, which are in general rather poorly constrained, have a tangible impact on the resulting \afe{}-\feh{} distribution of disc stars. This highlights the importance of quantifying these sources of systematic uncertainty, and of ensuring that models used to examine the evolution of galaxy elemental abundances are broadly compatible with orthogonal constraints. As discussed by \citet{2015MNRAS.446..521S}, the values of $\nu$ and $\tau$ adopted by EAGLE were calibrated to ensure that the simulations broadly reproduce the observed evolution of the cosmic Type Ia SNe rate density. 
\end{appendix}


\bsp	
\label{lastpage}
\end{document}